\documentclass[twocolumn]{aastex62}

\usepackage{xspace}

\newcommand{\nh}{\ensuremath{{N}_\mathrm{H}}\xspace}
\newcommand{\nhone}{\ensuremath{{N}_{\mathrm{H},1}}\xspace}
\newcommand{\nhtwo}{\ensuremath{{N}_{\mathrm{H},2}}\xspace}
\newcommand{\gx}{GX\,301$-$2\xspace}

\usepackage{amsmath}
\shorttitle{{\it X-Calibur} Observations of GX 301$-$2}
\shortauthors{{\it The X-Calibur team:} Abarr et al.}

\begin{document}

\title{Observations of a GX 301$-$2 Apastron Flare with the {\it X-Calibur}
Hard X-Ray Polarimeter\\ Supported by {\it NICER}, the {\it Swift} XRT and BAT, and {\it Fermi} GBM}
\author{Q.\,Abarr} \affil{Washington University in St. Louis, 1 Brookings Dr., CB 1105, St. Louis, MO 63130} 
\author{M.\,Baring} \affil{Rice University, Department of Physics \& Astronomy Department, 6100 Main Street, Houston TX 77251-1892} 
\author{B.\,Beheshtipour} \affil{
Max Planck Institute for Gravitational Physics (Albert Einstein Institute), Leibniz Universit\"at Hannover,
formely: Washington University in St. Louis, 1 Brookings Dr., CB 1105, St. Louis, MO 63130} 
\author{M.\,Beilicke} \affil{Former affiliation: Washington University in St. Louis, 1 Brookings Dr., CB 1105, St. Louis, MO 63130}
\author{G.\,de Geronimo} \affil{DG CIrcuits, 30 Pine Rd, Syosset, NY 11791}
\author{P.\,Dowkontt} \affil{Washington University in St. Louis, 1 Brookings Dr., CB 1105, St. Louis, MO 63130}
\author{M.\,Errando} \affil{Washington University in St. Louis, 1 Brookings Dr., CB 1105, St. Louis, MO 63130}
\author{V.\,Guarino} \affil{Guarino Engineering, 1134 S Scoville Ave  Oak Park, IL 60304} 
\author{N.\,Iyer} 
\affil{KTH Royal Institute of Technology, Department of Physics, 106 91 Stockholm, Sweden}
\affil{The Oskar Klein Centre for Cosmoparticle Physics, AlbaNova University Centre, 106 91 Stockholm, Sweden}
\author{F.\,Kislat} \affil{University of New Hampshire, Morse Hall, 8 College Rd, Durham, NH 03824} 
\author{M.\,Kiss} 
\affil{KTH Royal Institute of Technology, Department of Physics, 106 91 Stockholm, Sweden}
\affil{The Oskar Klein Centre for Cosmoparticle Physics, AlbaNova University Centre, 106 91 Stockholm, Sweden}
\author{T.\,Kitaguchi} \affil{RIKEN Nishina Center, 2-1 Hirosawa, Wako, Saitama 351-0198, Japan} 
\author{H.\,Krawczynski} \affil{Washington University in St. Louis, 1 Brookings Dr., CB 1105, St. Louis, MO 63130} 
\author{J.\,Lanzi} \affil{NASA Wallops Flight Facility, 32400 Fulton St, Wallops Island, VA 23337} 
\author{S.\,Li} \affil{Brookhaven National Laboratory, 98 Rochester St, Upton, NY 11973}
\author{L.\,Lisalda} \affil{Washington University in St. Louis, 1 Brookings Dr., CB 1105, St. Louis, MO 63130} 
\author{T.\,Okajima} \affil{NASA's Goddard Space Flight Center, Greenbelt, MD 20771}
\author{M.\,Pearce} 
\affil{KTH Royal Institute of Technology, Department of Physics, 106 91 Stockholm, Sweden}
\affil{The Oskar Klein Centre for Cosmoparticle Physics, AlbaNova University Centre, 106 91 Stockholm, Sweden}
\author{L.\,Press} \affil{Washington University in St. Louis, 1 Brookings Dr., CB 1105, St. Louis, MO 63130} 
\author{B.\,Rauch} \affil{Washington University in St. Louis, 1 Brookings Dr., CB 1105, St. Louis, MO 63130} 
\author{D.\,Stuchlik} \affil{NASA Wallops Flight Facility, 32400 Fulton St, Wallops Island, VA 23337} 
\author{H.\,Takahashi} \affil{
Hiroshima University, Department of Physical Science, 1-3-1, Kagamiyama, Higashi-Hiroshima, 739-8526, Japan} 
\author{J.\,Tang} \affil{Washington University in St. Louis, 1 Brookings Dr., CB 1105, St. Louis, MO 63130} 
\author{N.\,Uchida} \affil{
Hiroshima University, Department of Physical Science, 1-3-1, Kagamiyama, Higashi-Hiroshima, 739-8526, Japan} 
\author{A.\,West\newline} \affil{Washington University in St. Louis, 1 Brookings Dr., CB 1105, St. Louis, MO 63130} 

\author{P.\,Jenke} \affil{University of Alabama in Huntsville, Huntsville, AL 35899, USA} 
\author{H.\,Krimm} \affil{National Science Foundation, 2415 Eisenhower Ave., Alexandria, VA 22314  USA} 
\author{A.\,Lien} 
\affil{Center for Research and Exploration in Space Science and Technology (CRESST) and NASA Goddard Space Flight Center, Greenbelt, MD 20771, USA}
\affil{Department of Physics, University of Maryland, Baltimore County, 1000 Hilltop Circle, Baltimore, MD 21250, USA}
\author{C.\,Malacaria} 
\affil{ST12 Astrophysics Branch, NASA Marshall Space Flight Center, Huntsville, AL 35812, USA}
\affil{Universities Space Research Association, NSSTC, Huntsville, AL 35805, USA}
\author{J.\,M.\,Miller} \affil{University of Michigan, Department of Astronomy, 1085 S. University, Ann Arbor, MI 48109} 
\author{C. Wilson-Hodge} \affil{ST12 Astrophysics Branch, NASA Marshall Space Flight Center, Huntsville, AL 35812, USA}
\correspondingauthor{Henric Krawczynski, krawcz@wustl.edu,
Fabian Kisklat, fabian.kislat@unh.edu, and Manel Errando, errando@wustl.edu}

\begin{abstract}
The accretion-powered X-ray pulsar GX\,301$-$2 was observed
with the balloon-borne {\it X-Calibur} hard X-ray 
polarimeter during late December 2018, with contiguous observations by the {\it NICER} X-ray telescope, the {\it Swift} X-ray Telescope and Burst Alert Telescope, and the {\it Fermi} Gamma-ray Burst Monitor spanning several months.  The observations detected the pulsar in a rare apastron flaring state coinciding with a significant spin-up of the pulsar discovered with the {\it Fermi} GBM. 
The {\it X-Calibur}, {\it NICER}, and {\it Swift} observations reveal a pulse profile strongly dominated by one main peak, and the {\it NICER} and {\it Swift} data show strong variation of the profile 
from pulse to pulse. The {\it X-Calibur} observations constrain for the first time the linear polarization of the 15-35\,keV emission from a highly magnetized accreting neutron star, indicating a polarization degree of $(27^{+38}_{-27})$\% (90\% confidence limit) averaged over all pulse phases.
We discuss the spin-up and the X-ray spectral and polarimetric results in the context of theoretical predictions. We conclude with a discussion of the scientific potential of future observations
of highly magnetized neutron stars with the more sensitive follow-up mission {\it XL-Calibur}.
\end{abstract}
\hspace*{2cm}\\[2ex]
\section{Introduction}
\label{sec:intro}
\begin{figure}[t]
\plotone{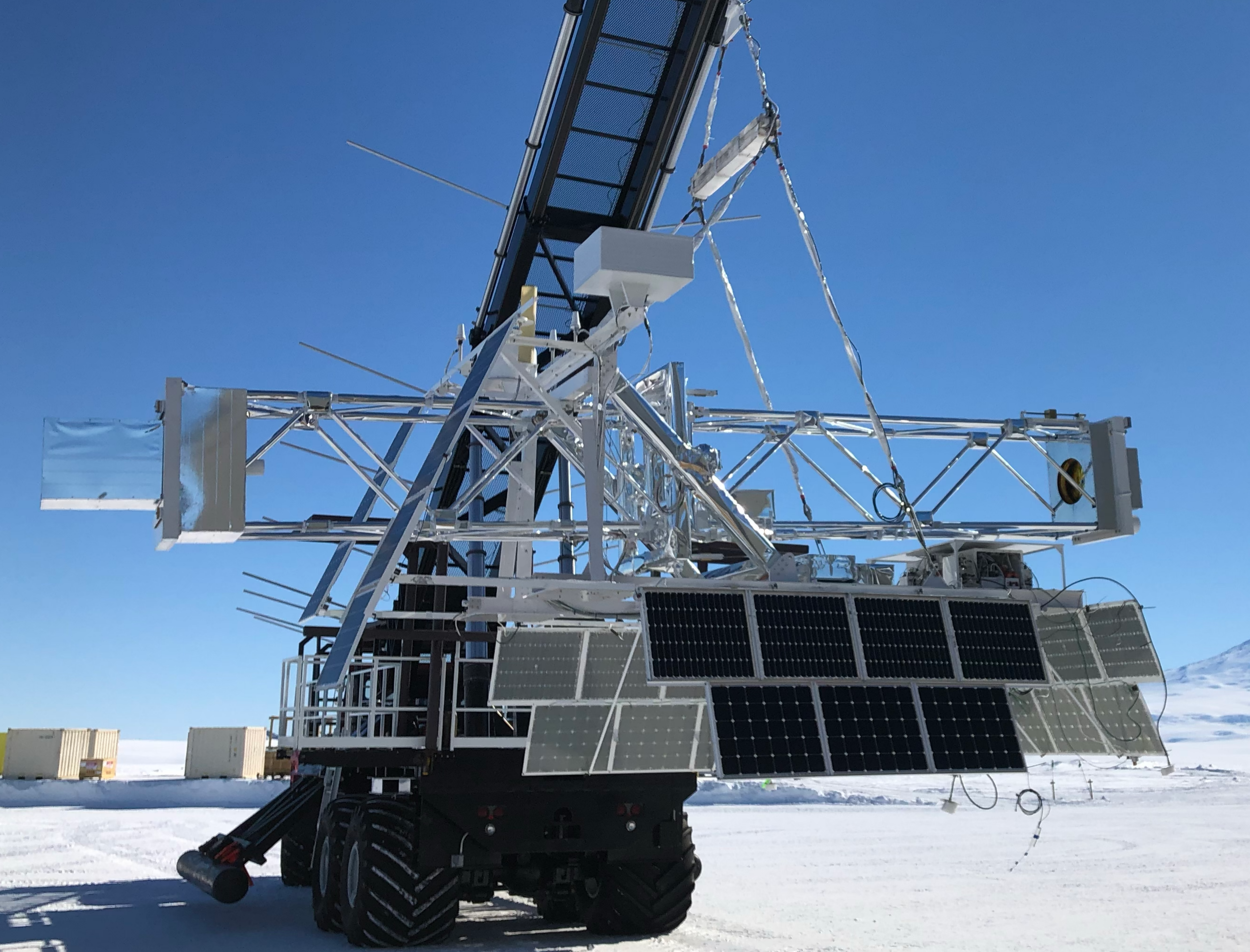}
\caption{\label{f:xcal} The {\it X-Calibur} hard X-ray polarimeter during
integration in McMurdo (Antarctica) in December 2018. The {\it InFOC$\mu$S}
X-ray mirror is used to focus the X-rays onto a scattering polarimeter at the front (right) and back (left) ends of the 8\,m long telescope.}
\label{fig:X-Calibur_McMurdo}
\end{figure}

In this paper, we report on phase-resolved spectro-polarimetric observations of the accretion-powered, highly-magnetized X-ray pulsar GX 301$-$2 
with the {\it X-Calibur} baloon-borne mission (see Fig.\ \ref{f:xcal}) \citep{Kraw:11a,Guo:13,Beil:14,Beil:15,Kisl:17,Kisl:18} in late December 2018.
The observations were accompanied by 
spectro-temporal observations in overlapping and adjacent periods by the
{\it Neil Gehrels Swift Observatory (Swift)}
Burst Alert Telescope (BAT) \citep[]{Bart:05}, the {\it Swift} X-ray Telescope (XRT) \citep{Burr:07}, the {\it Neutron star Interior Composition Explorer Mission (NICER)} X-ray telescope \citep{Gend:12}, and the {\it Fermi} Gamma-ray Burst Monitor (GBM) \citep{Meeg:09}.
The observations covered a particularly interesting epoch in which the pulsar exhibited rare flaring activity associated with a substantial pulsar spin-up. 

The pulsar is in an orbit of period $\sim$41.5 days and eccentricity 0.462 about the star Wray 977, also known as BP Crucis \citep{Koh:97,Sato:86,Doro:10}, an extremely bright B1\,Ia\,Hypergiant at a distance of $4.0^{+0.6}_{-0.5}$\,kpc \citep{2018yCat.1345....0G}. 
Wray~977 has an estimated mass of $\sim$39-63 M$_{\odot}$, a radius of $\sim 60 R_{\odot} \sim 0.3$AU, 
and shines with a bolometric luminosity of  $\sim$5$\times 10^5$ $L_{\odot}$ \citep{Kape:06,Clar:12}.
The pulsar has a spin period of $\sim$680\,sec  \citep{Whit:76} and a 2-10\,keV luminosity of 
$10^{37}$-$10^{38}$ erg/s \citep{Liu:18}. 
The pulsar displays bright flares prior to periastron at an orbital phase of ${\sim}0.93$ \citep{Leah:08}.  Although GX 301$-$2 is in a tight orbit with a 
hypergiant star (its semi-major axis is $\sim 19-30$AU), the X-ray light curves do not show any evidence for eclipses.
\citet{Park:80}, \citet{Kape:06}, \citet{Leah:08}   
estimate the inclination (angle between the binary angular momentum vector 
and the observer) to lie between 44$^{\circ}$ and 78$^{\circ}$.   
\begin{figure*}[ht]
\centerline{\includegraphics[width=14cm]{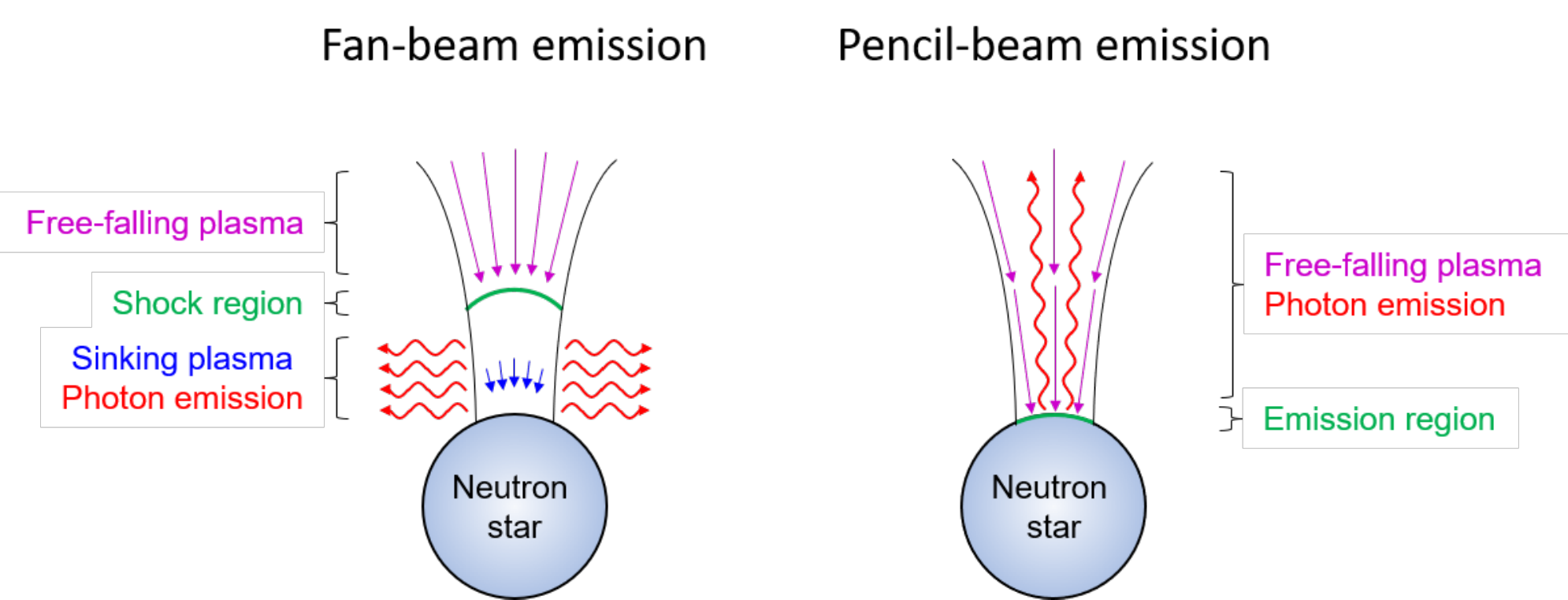}}
\caption{\label{f:emission} Schematic views of the fan-beam (left) and pencil-beam (right) emission geometries. (Adapted from\,\citet{2007A&A...472..353S}.)}
\end{figure*}

GX 301$-$2 is believed to accrete from the wind of its companion, and possibly from a plasma stream \citep{Leah:08}
or a temporary accretion disk \citep{Koh:97,Nabi:19}. As the material sinks toward the neutron star, 
it latches onto the magnetic field lines at the magnetospheric radius $r_{\it m} \sim 2000-3000\;$km
from the center of the neutron star \citep[][and references therein]{Lipu:92,Mesz:92}.
Transferring its angular momentum to the neutron star, the plasma moves along the 
magnetic field lines until it dissipates its kinetic energy either in a radiative 
shock above the neutron star surface or in a hydrodynamic shock right at the neutron star surface
\citep{Bask:75,Bask:76,Beck:12,Mush:15a}. 
The X-ray emission is believed to form through the Comptonization of black body, bremstrahlung, and cyclotron seed photons emitted in and nearby the shocked plasma leading to a power law at low energies with an exponential cutoff in the \mbox{10-20\,keV} energy range \citep{Beck:07,Fari:12,Post:15,West:17,Wolf:19}.

The literature on accreting X-ray pulsars distinguishes between two idealized radiation patterns
associated with the different locales for the energy dissipation, as illustrated in Fig.\,\ref{f:emission}.  
The dissipation in a radiative shock further up in the accretion column
is believed to lead to a fan-shaped radiation pattern with most photons
leaving the accretion column perpendicular to the flow direction \citep{Davi:73}.
Emission associated with a hydrodymnamic shock close to the neutron star surface is expected to lead to a more narrowly focussed emission pattern
resembling a pencil beam \citep{Burn:91,Nels:93}.  
Discriminating between these two scenarios is a prime goal of studies of X-ray Binaries, and X-ray polarimetry stands to play a decisive role.

GX 301$-$2 observations with {\it NuSTAR} revealed two 
cyclotron resonant scattering features (CRSFs)
with line centroids $E_{\rm CRSF}$ 
and Gaussian widths $\sigma_{\rm CRSF}$ of $(E_{\rm CRSF},\sigma_{\rm CRSF})$= (37\,keV, 5\,keV)
and (50\,keV, 8\,keV) \citep{Fuer:18,Nabi:19}. The CRSF energies and widths depend on time and on the pulsar phase \citep{Krey:04,Fuer:18,Nabi:19}. 
In XRBs, CRSFs are associated with electrons transitioning between quantized Landau levels, 
the transverse energy discretization relative to the magnetic field direction that emerges from the Dirac equation in quantum electrodynamics (QED).
The observation of an \underline{electron} CRSF at energy $E_{\rm CRSF}$ constrains the magnetic field to be:
\begin{equation}
   B\, \approx\, \frac{(1+z)}{n} 
   \frac{E_{\rm CRSF}}{11.57\,\rm\,keV}\,10^{12}\rm \,\,G.
 \label{eq:B_CRSF}
\end{equation}
Here, the positive integer $n$ is the harmonic number of the cyclotron transition.  This relation applies to 
line features at energies significantly lower than
$m_ec^2$, i.e.\,when $B$ is much smaller than the quantum critical field $B_{\rm cr}\,=\,\frac{m^2 c^3}{e \hbar}\;\approx\; 4.41 \times 10^{13}\,\,G$, so that the harmonics are evenly spaced.

For a neutron star of mass $M$ and an emission from radius $r_{\rm em}$ 
(measured from the center of the neutron star),
the redshift $z$ is approximately given by:
\begin{equation}
z\,=\,\frac{1}{\sqrt{1-\frac{2 G M}{r_{\rm em} c^2}}}-1\,\approx\,
0.15 \,\frac{M}{M_{\odot}} \left(\frac{r_{\rm em}}{10\,\rm km}\right)^{-1}.
\end{equation}
If the absorption features are interpreted as coming from one region, then the natural $n=3,4$ inference would yield $B=1.1 \times 10^{12} (1+z)\,$Gauss, while an $n=2,3$ choice gives $B\sim 1.5 \times 10^{12} (1+z)\,$Gauss.  In such a case, the absence of a prominent $n=1$ fundamental at lower energies poses an issue.  Thus, \cite{Fuer:18} interpret the two features as being fundamentals from distinct regions, in which case they 
possess higher fields, namely $\sim 3\times 10^{12}$G and $\sim 4.3\times 10^{12}$G (for $z=0$), corresponding to cyclotron absorption radii differing by only around 12\%.  These fields are substantially above the values inferred from accretion torque models \citep[see Table\,1 of][]{Stau:19}, the converse of what is usually obtained when comparing these two field estimates for X-ray binary pulsars.
Some CRSFs are observed to depend on pulse phase, time, and luminosity \citep[][and references therein]{Stau:19}. 
These variations are sometimes attributed to a movement of the 
radiative shock along the accretion column, or by changes 
in the magnetic field geometry \citep[e.g.][]{Beck:12,Mush:15b}.

The polarimetric capability of {\it X-Calibur} opens up a new degree of freedom in diagnosing the physical environment of GX 301$-$2.
Observations of the linear polarization fraction and angle can provide 
qualitatively new information on the origin of X rays in the accretion column or at its impact locale on the neutron star surface, on their birefringent propagation in the magnetosphere, and on the photon interaction cross sections. 

The predictions of the polarization of the X-rays from
highly magnetized neutron stars
depend strongly on the strong-field 
Quantum Electrodynamic (QED) 
predictions of the birefringence of the magnetized vacuum
\cite{Eule:35,Heis:36,Weis:36,Schw:51,Toll:52,Gned:74,Chan:79,Heyl:00} 
and the mode dependence of the scattering cross sections
and absorption coefficients \citep[e.g.][]{Adle:70,Canu:71,Adle:71,Mesz:78,Vent:79,Aron:87,Mesz:92,Hard:06}.
\citet{Kii:86,Kii:87} and \citet{Mesz:88} used polarization-dependent radiation transfer 
calculations to predict the polarization 
fractions of accreting X-ray pulsars. 
They found that the mode-dependent scattering cross-sections lead to high polarization fractions 
in certain pulse intervals. 
\citet{Mesz:88} determined that the models robustly predict that the phase-resolved 
flux and polarization fraction should be correlated (anti-correlated) in the fan beam
(pencil beam) models. The detection of such a correlations can therefore discriminate between the fan beam and pencil beam models. This is a design driver for an upgraded version of {\it X-Calibur}, as described in Sect.\,\ref{sec:discussion}.
 
The rest of the paper is structured as follows.  
The {\it X-Calibur} mission and experiment is described in Sect.\,\ref{sec:xcal}.  
The {\it X-Calibur}, {\it NICER}, {\it Swift},
and {\it Fermi} observations and data analysis methods are described in Sects.\,\ref{sec:obs} and \ref{sec:ana}, respectively.
We present the results of the observations in  Sect.\,\ref{sec:results} and conclude with a summary and an outlook for 
the scientific potential of 
follow-up flights in Sect.\,\ref{sec:discussion}.  
The appendices include a description of the {\it X-Calibur} Stokes parameter analysis (Appendix A),
our estimates of the systematic errors on the {\it X-Calibur} polarization parameters (Appendix B), and a summary
of the spectral results (Appendix C).

All errors and uncertainties are quoted at 1$\sigma$-level (68.27\% confidence level), unless noted otherwise.
\section{The {\it X-Calibur} Experiment} 
\label{sec:xcal}
\begin{figure}[t]
\plotone{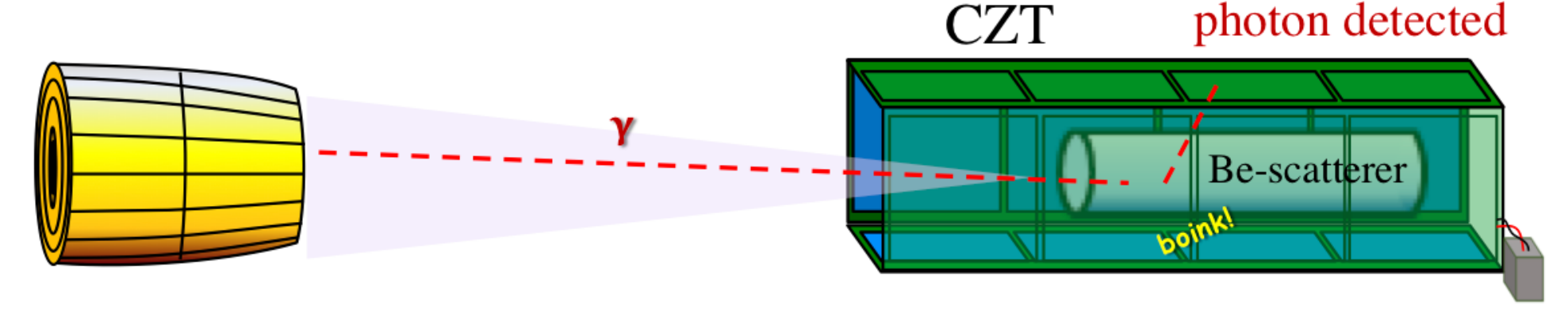}
\caption{\label{f:dp} {\it X-Calibur} detection principle: the X-ray mirror focuses photons onto a 
Be scattering element. The scattered photon is detected in the surrounding assembly of CZT detectors.
The distribution of the azimuthal scattering angles depends on the linear polarization fraction and angle.
A rear CZT detector behind the scattering element (at
the right side of the detector assembly) is used to
monitor the position of the source in the field of view.}
\end{figure}

{\it X-Calibur} combines an 8\,m long X-ray telescope 
with arc-second pointing and a scattering polarimeter (Fig.\,\ref{fig:X-Calibur_McMurdo}). 
The telescope uses an aluminum-carbon fiber 
optical bench \citep{Kisl:17}, which is pointed 
with the Wallops Arc Second Pointer (WASP)
with a pointing stability of $\sim$1'' 
Root Mean Square and a pointing 
knowledge of $<$15'' (3$\sigma$) \citep{Stuc:15}.
{\it X-Calibur's} energy range is limited to $>$15~keV
by the absorption in the residual atmosphere at a 
float altitude of 125,000 feet, and to $<$60 keV by
the mirror reflectivity.
The mirror achieves an angular resolution 
of 2.5~arcmin Half-Power Diameter  
and effective areas of 93 cm$^2$ at 20\,keV and 46 cm$^2$ at 35\,keV \citep{Okaj:02,Bere:03,Tuel:05,Ogas:08}.
Grazing incidence mirrors 
reduce the polarization of cosmic X-ray signals 
by less than 1\% of the true polarization 
owing to the shallow scattering angles \citep{Sanc:93,Kats:09}.
The polarimeter is shown in Fig.\ \ref{f:dp} and 
is made of a Be scattering element 
inside an assembly of Cadmium Zinc Telluride (CZT) detectors (each 2~mm thick, 2$\times$2\,cm$^2$ 
footprint, 64 pixels).
Photons 
preferentially scatter  perpendicular to the angle of the 
electric field of the beam with
an azimuthal scattering angle distribution of: 
\begin{equation}
\frac{dN}{d\psi} \propto \frac{1}{2\pi} \left[1+ \mu\, p_0 \,\cos{(2 (\psi-\psi_0 - \pi/2))}\right],
\label{e:mu}
\end{equation}
with $p_0$ and $\psi_0$ being the true polarization fraction and angle, $\psi$ the measured azimuthal scattering angle, and $\mu=51.3\%$ is 
{\it X-Calibur's} modulation factor. 
A rear CZT detector is positioned behind 
the scattering slab for monitoring 
the source location in the field-of-view. 
The timing resolution is \mbox{$\sim 1\,\mu$s}. The
energy resolution increases from $\sim$3\,keV FWHM at
15\,keV to 5\,keV FWHM at 35\,keV.
The detector assembly is shielded by a fully active CsI(Na) shield, and the polarimeter/shield 
assembly rotates at 1\,rpm around the optical 
axis to minimize systematic errors.
Detailed descriptions of the polarimeter and the 
in-flight performance of all components are given in \citep{Beil:14,Kisl:18,Abar:19a}. 
\section{Observations}  
\label{sec:obs}
{\it X-Calibur} was launched at 20:45 on Dec.\,29, 2018 (all times and dates are UTC) 
and reached a float altitude of 39.6 km (130,000 feet) roughly 3 hours later. 
Following the checkout of the pointing system and the in-flight optimization 
of the anti-coincidence shield settings, {\it X-Calibur} observed the accreting X-ray pulsars 
GX\,301-2 and Vela\,X-1 until the flight was aborted
owing to a He leak of the balloon at 10\,pm on Jan.\,1, 2019. The starting times and durations of the {\it X-Calibur}
on-source observations windows  
are listed in Table \ref{tabO}. 
The high-balloon-altitude GX\,301$-$2 data set comprises a total of 8.0 hours ON-source and 
7.8 hours OFF-source (aiming $1^{\circ}$ away from the source).

The {\it NICER} X-ray Timing Instrument \citep[XTI, ][]{2016SPIE.9905E..1HG} observed \gx\ for 2.0\,ksec on 2018 Dec.\ 28 and for 0.2\,ksec on Dec.\ 29. Observations on Dec.\ 28 were split into five shorter observation windows and were analyzed independently.
The {\it Swift} X-ray Telescope (XRT) observed GX 301-2 from MJD\,58,480 through MJD\,58,488 in nine individual pointings between 0.5\,ks and 1.1\,ks for a total of 8.1\,ks. Details about the {\it NICER} and {\it Swift} observations are summarized in Appendix C, Table \ref{tabO}. { Observing windows are labeled 
{\it X-I - X-XXXIV} for {\it X-Calibur},
$N-I$ to $N-V$ for {\it NICER} and $S-I$ to $S-IX$ for {\it Swift}-XRT.}
The {\it Swift} BAT and {\it Fermi} GBM observe GX 301$-$2 on a regular basis. 
We use in the following results obtained for individual orbits and results averaged over individual days.  \newpage

\section{Data Analysis}
\label{sec:ana}
\subsection{{\it X-Calibur} Data Analysis}
The {\it X-Calibur} data analysis uses single-pixel CZT events without shield veto.  
The energy deposited in the CZT is estimated based on the calibration of the polarimeter 
with a $^{152}$Eu source with low-energy lines at 39.52\,keV (K$_{\alpha 2}$), 
40.12\,keV (K$_{\alpha 1}$), 45.7\,keV, and 121.78\,keV. 

An event consists of the pixel number $i$ (located at position $\vec{x}_i=(x,y,z)_i$ 
in the detector reference frame with $x$ and $y$ being the coordinates in the focal plane 
and $z$ pointing towards the source), the energy $E$ deposited in the CZT detectors, 
and the GPS event time\,$t$.  Consistent with the exponential cutoff of the energy 
spectrum \citep[e.g.][]{Fuer:18}, {\it X-Calibur} does not detect a significant excess 
of photons with $>$35\,keV energy deposits, and we thus only use $E<35$\,keV events.
{ The events enter the analysis with weights that were optimized 
based on the detector response as inferred from Monte Carlo simulations  (Appendix A).} 
For light curves, we normalize the weights so that the weighted event rate equals the true source rate.

The polarization analysis uses the Stokes parameters $I$ (total flux), 
$Q$ (the linearly polarized flux along the North-South direction), and 
$U$ (the linearly polarized flux along the direction rotated 45$^{\circ}$ 
counterclockwise from the North-South direction 
when looking at the source)
which are the weighted sums of the corresponding Stokes parameters of individual events
\citep{Kisl:15,Stro:17}.
The main results are given in terms of the normalized Stokes parameters: 
\begin{align}
  {\cal Q}  & = Q/I\\
  {\cal U}  & = U/I
\end{align}
so that ${\cal Q}$ (${\cal U}$) equals 1 for a beam 100\% linearly polarized along the 
North-South (Northeast-Southwest) direction.
The reconstructed polarization fraction $p_{\rm r}$ is given by:
\begin{equation}
p_{\rm r}\,=\,\sqrt{{\cal Q}^2+{\cal U}^2}
\end{equation}
and the reconstructed polarization angle $\psi_{\rm r}$ is given by:
\begin{equation}
\psi_{\rm r}\,=\,\frac{1}{2}\,\arctan{({\cal U/Q})}\, =\,\frac{1}{2}\,\arctan{(U/Q)}.  
\end{equation}
During the observations, we switch every 15 minutes between observations targeting  GX 301$-$2 (ON observations) and observations 
of four fields each located in a cross-pattern 1$^{\circ}$ away from the source in pitch and in yaw (OFF observations). 
As the Stokes parameters are additive, we can infer the Stokes parameters of the source beam 
by calculating the Stokes parameters for the ON-observations and OFF observations, and 
subtracting the OFF values from the ON values after scaling the OFF values according 
to the ON and OFF observation time ratio. Details of the Stokes parameter analysis and background 
subtraction procedure are given in Appendix A.
The systematic error on a measured polarization fraction $p_{\rm r}$ is (Appendix B):
\begin{equation}
\label{e:se}
\Delta p_{\rm r}\,=7.25\%\times p_{\rm r}.
\end{equation}
The error $\Delta p_{\rm r}$ is our best 
estimate of the maximum possible error. 

We fit the {\it X-Calibur} energy spectrum with {\it XSPEC} \citep{Arna:96,Arna:18} using
Response Matrix Files (RMFs) and Auxiliary Response Files (ARFs) derived from Monte Carlo simulations.

\subsection{{\it NICER}, {\it Swift}, and {\it Fermi} Data Analysis}  
The {\it NICER} data were processed using \texttt{NICERDAS v2018-11-19\_V005a} included in \texttt{HEASOFT v6.25}. Data were calibrated, cleaned, and combined using the \texttt{nicerl2} script with default screening filters. For spectral analysis, channels corresponding to  energies \mbox{2-10\,keV} were selected. 

The {\it Swift} XRT data were taken entirely in windowed timing mode analyzed with the CALDB version 20180710 and with \texttt{HEASOFT v6.25}, using {\texttt swxwt0to2s6\_20131212v015} response function. The absorption models were fit within the {\it xspec} command. 

The {\it Swift} BAT data analysis uses the \texttt{HEASOFTv6.23} software and BAT CALDB version 20171016. The BAT light curves in eight energy bands (14-20, 20-24, 24-35, 35-50, 50-75, 75-100, 100-150, and 150-195 keV) are created from the BAT survey data 
with the same methodology that was used for the previous BAT survey catalogs \citep{Oh:18,Baum:13}.
The 15-50 keV light curve is from the BAT transient 
monitor \citep{Krim:13}.

The {\it Fermi} GBM results were
taken from the National Space, Science, and Technology Center (NSSTC) web page.
The results are derived from the GBM NaI detectors binned in 0.256 s time bins and use the 12-25\,keV and 25-50\,keV energy channels \citep{Conn:19}.
The spin-frequencies are extracted using techniques described in \citep{Fing:99,Jenk:12}.

{
\subsection{Orbital and Pulsar Phases}
We compute the orbital phase with the parameters from \citet{Doro:10}, with the last recorded periastron passage on 
MJD 53531.65$\pm$0.01, an orbital period of
$P\,=$\,41.472 days, 
and a period derivative of
$\dot{P}\,=$\,$(-3.7\pm0.5)\times10^{-6}$ sec/sec.

We calculate the pulsar phase with the following 
phase model derived from {\it Fermi}-GBM data:
\begin{equation}
    \phi(t)\,=\,\dot{\phi} (t-t_0)
    +\frac{\ddot{\phi}}{2}(t-t_0)^2
    +\frac{\dddot{\phi}}{6}(t-t_0)^3
    +\frac{\ddddot{\phi}}{24}(t-t_0)^4
\end{equation}
with $t$ being the barycentered time. 
The model parameters are given in Table \ref{t:eph}.} 
\section{Results}  
\label{sec:results}

\begin{figure}[t]
\epsscale{1.25}
\plotone{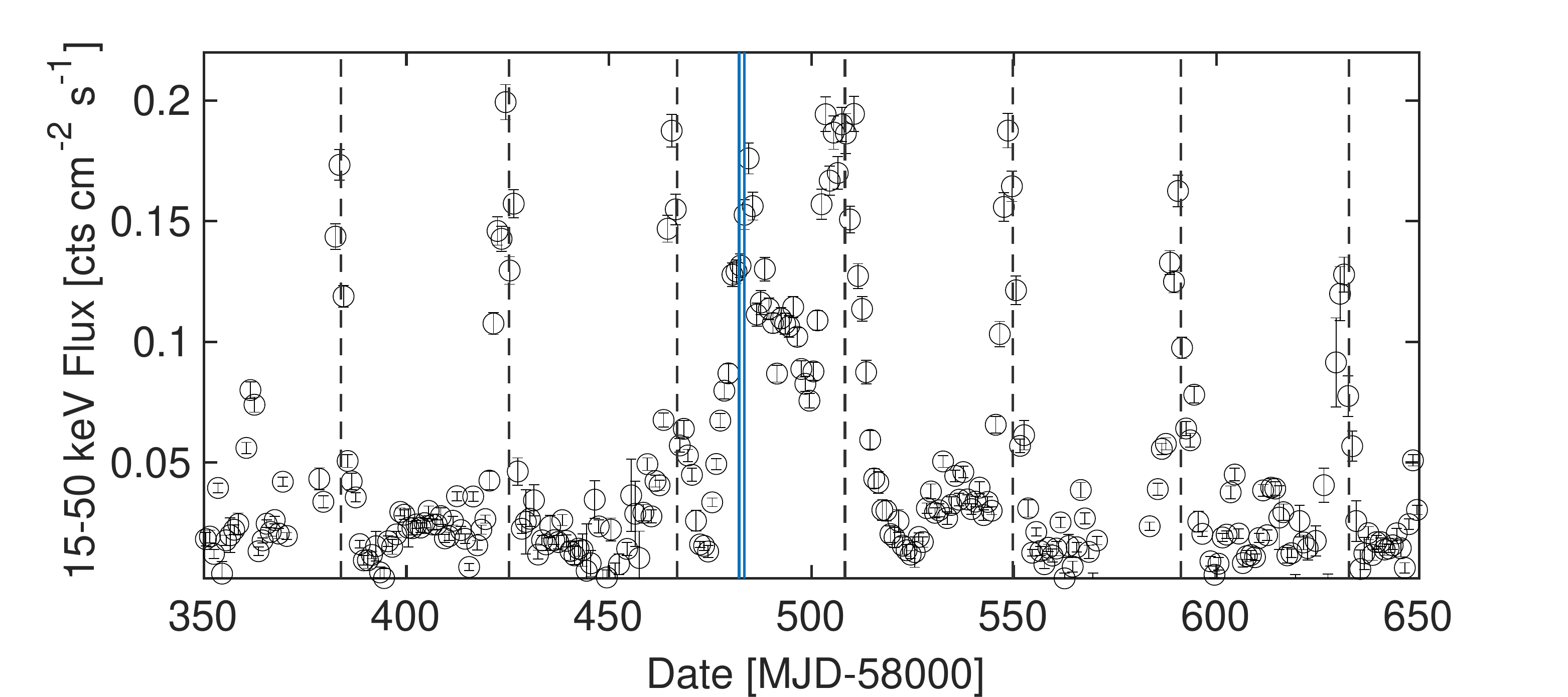}
\caption{\label{f:bat1} GX 301$-$2 15-50\,keV fluxes measured with the {\it Swift} BAT instrument \citep{Lien:2019}. The time interval of the
{\it X-Calibur} observations is marked by the solid blue vertical lines. The periastron passages are marked by the
dashed black vertical lines. 
}
\end{figure}
\begin{figure}[t]
\epsscale{1.25}
\plotone{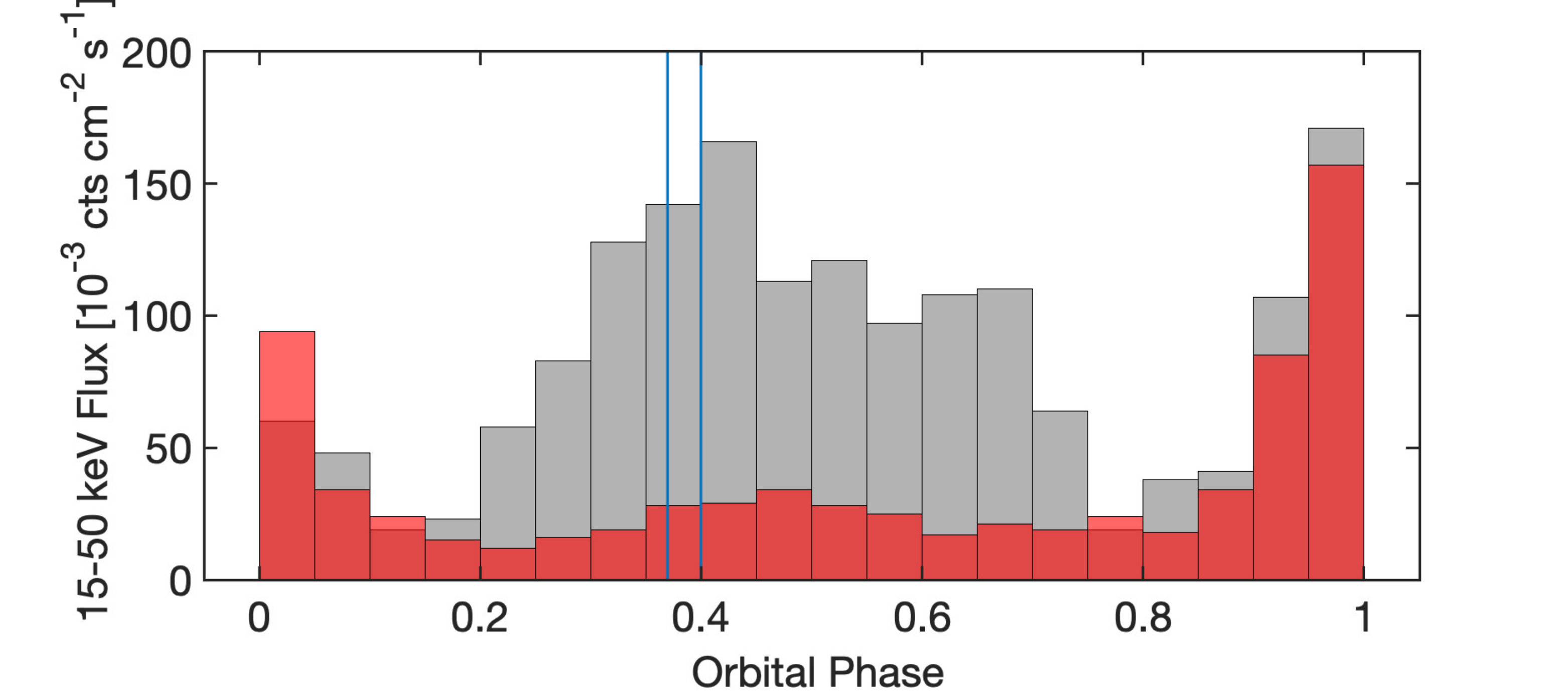}
\caption{\label{f:bat2} Average GX 301$-$2 {\it Swift} BAT 15-50\,keV flux 
in the 11 orbital cycles before the apastron flare (red
histogram), and in the orbital cycle of the apastron 
flare (grey histogram). The time interval of the
{\it X-Calibur} observations is marked by two vertical blue lines.}
\end{figure}

{ 
\subsection{Timing Results}}
Figure \ref{f:bat1} shows the 15-50\,keV fluxes measured with the Swift BAT. 
The graph clearly shows the 41.5\,day orbital period. The {\it X-Calibur} observations from \mbox{MJD 58,482.1521-58,483.3912} { (orbital phases 0.37-0.40)} fall into a rare period of a flare close to apastron. 
Figure\,\ref{f:bat2} compares the {\it Swift} BAT 15-50\,keV count rate 
measured during the orbit covering the apastron flare with the average 
count rates measured during the previous eleven orbits.  The activity was enhanced during the orbit of the apastron flare, 
with a pronounced peak at an orbital phase around 0.4.

\begin{figure}[t]
\epsscale{1.25}
\plotone{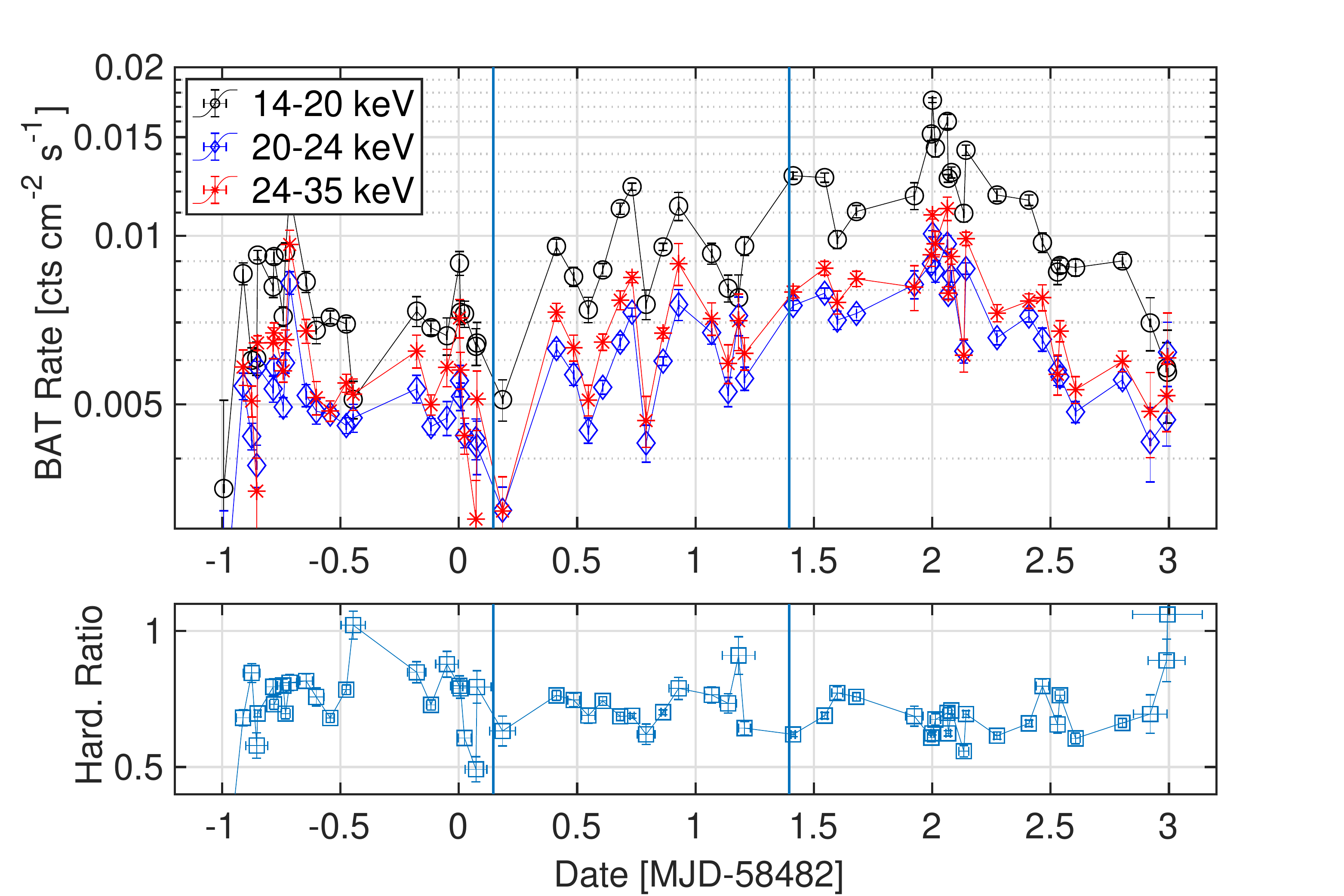}
\caption{\label{f:bat3} GX 301$-$2 fluxes measured 
with the Swift BAT instrument in three different energy bands (top panel) and the 24-35\,keV to 14-20\,keV hardness ratio 
(bottom panel) during the apastron flare.
The time interval of the {\it X-Calibur} observations 
is marked by two vertical blue lines.}
\end{figure}

The {\it Swift} BAT data allows us to scrutinize
the hard X-ray emission for spectral variability.
Figure\,\ref{f:bat3} presents the 
14-20\,keV, 20-24\,keV, and 24-35\,keV light curves
and the 24-35\,keV to 14-20\,keV hardness ratios. The RMS of the
hardness ratios is 0.103 corresponding to a RMS of the photon indices $\Gamma$ (from $dN/dE\propto E^{-\Gamma}$) of 
$\Delta \Gamma\,\approx$\,1. We do not discern a clear pattern linking the hardness ratio
excursions to the flux level or the flux history except for a pronounced hardening 
of the energy spectra at the end of the flaring periods at MJD\,58,481.5539, MJD\,58,483.1819 and MJD\,58,484.9899.

\begin{figure}[t]
\epsscale{1.25}
\plotone{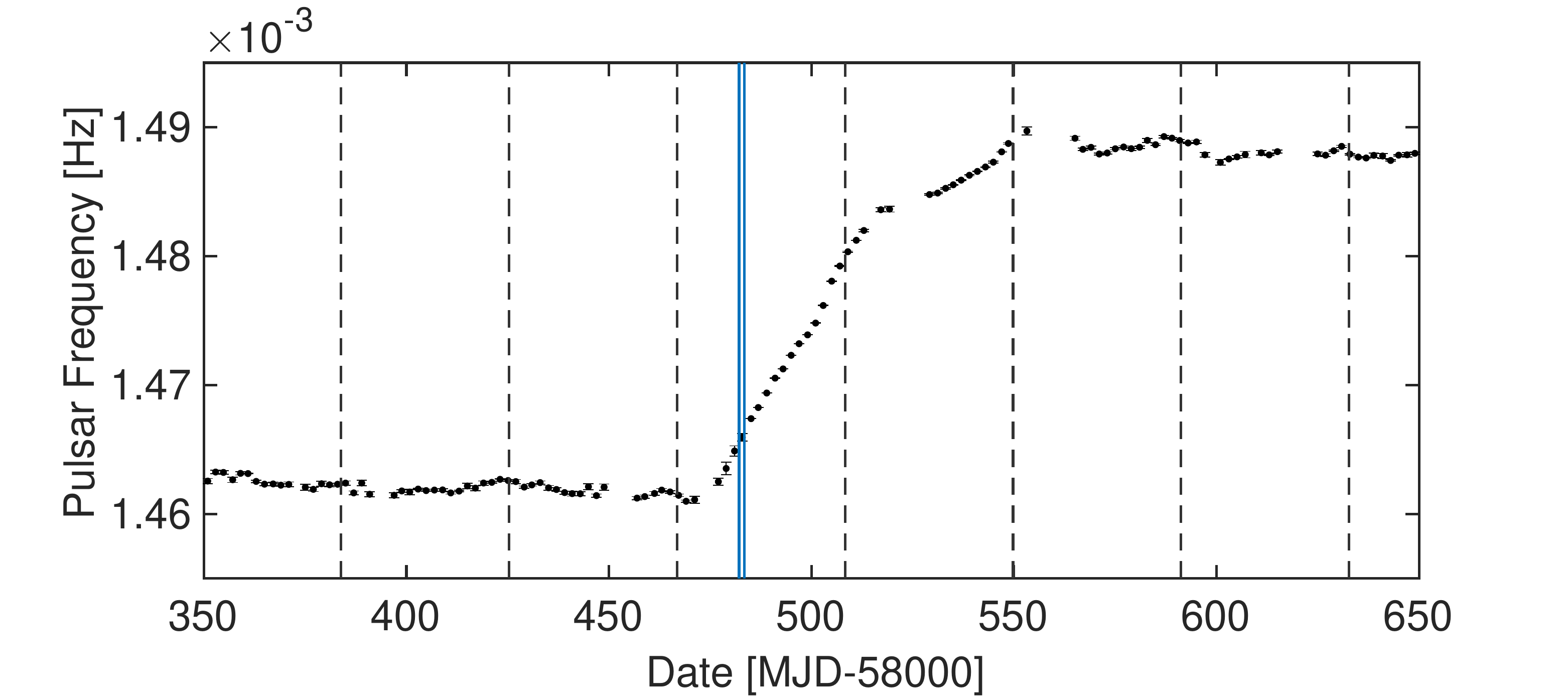}
\caption{GX 301$-$2 spin frequency as measured by the {\it Fermi} GBM from \mbox{12-50\,keV} observations between $20^{\rm th}$ August 2018 and $16^{\rm th}$ June 2019
\citep[from the][]{Conn:19}. The time period of the {\it X-Calibur} observations is shown by two vertical blue solid lines. The periastron passages are marked by vertical dashed lines.
}
\label{f:gbm}
\end{figure}

\begin{figure}[t]
\epsscale{1.25}
\plotone{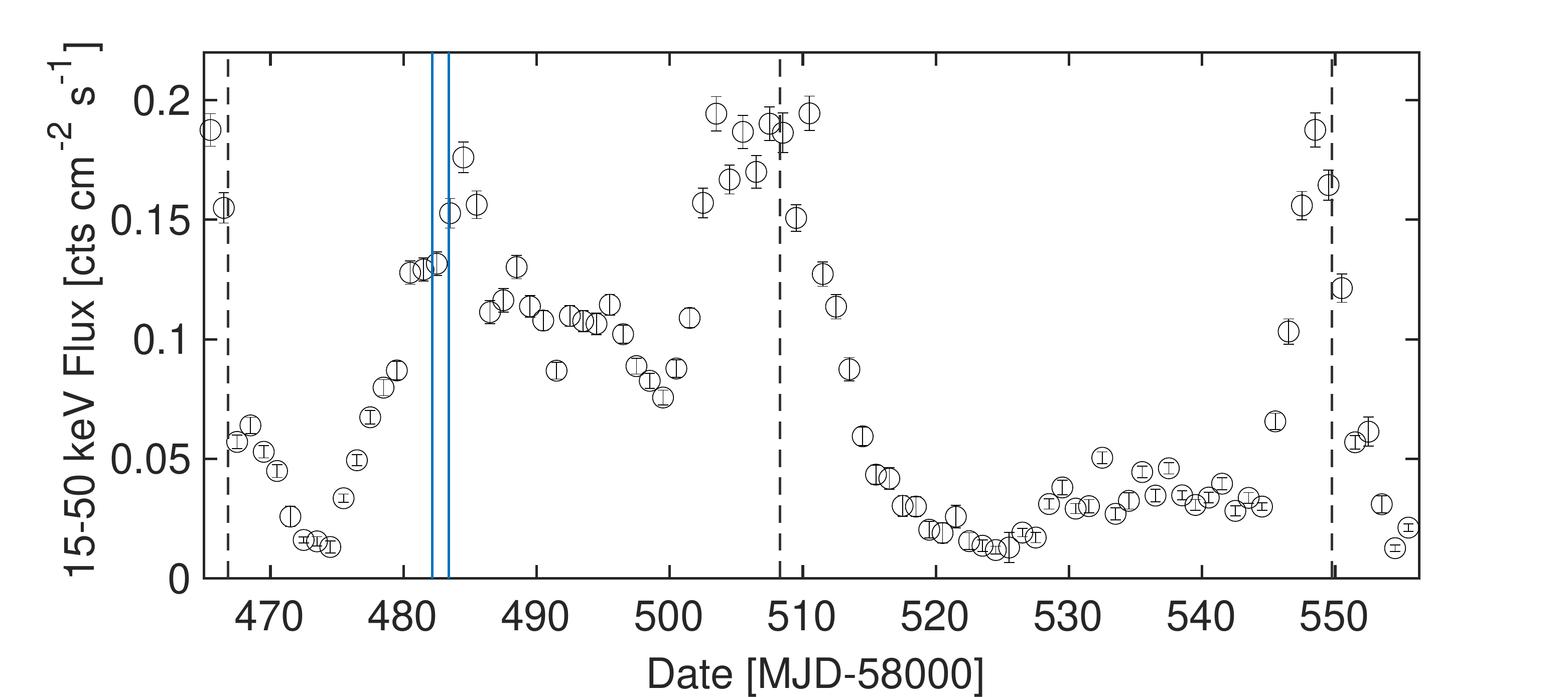}
\plotone{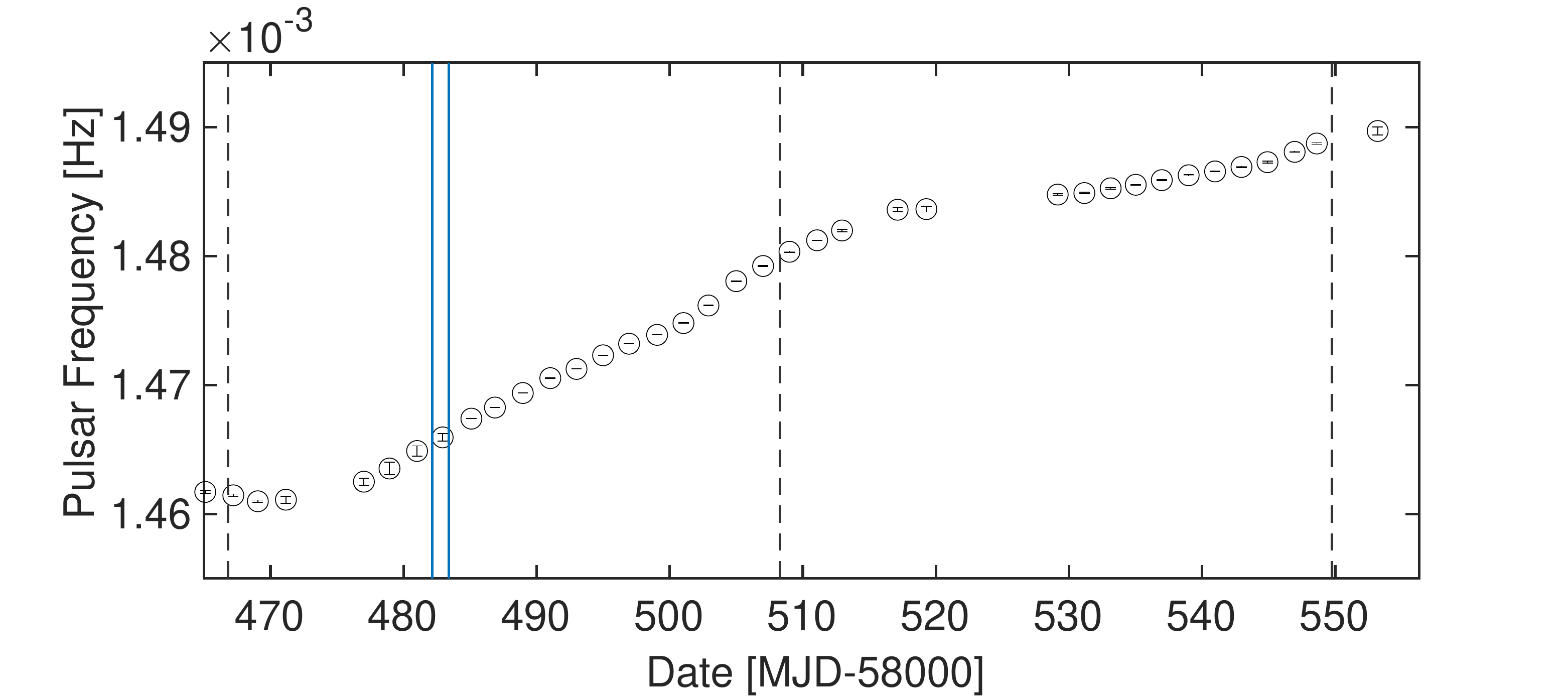}
\plotone{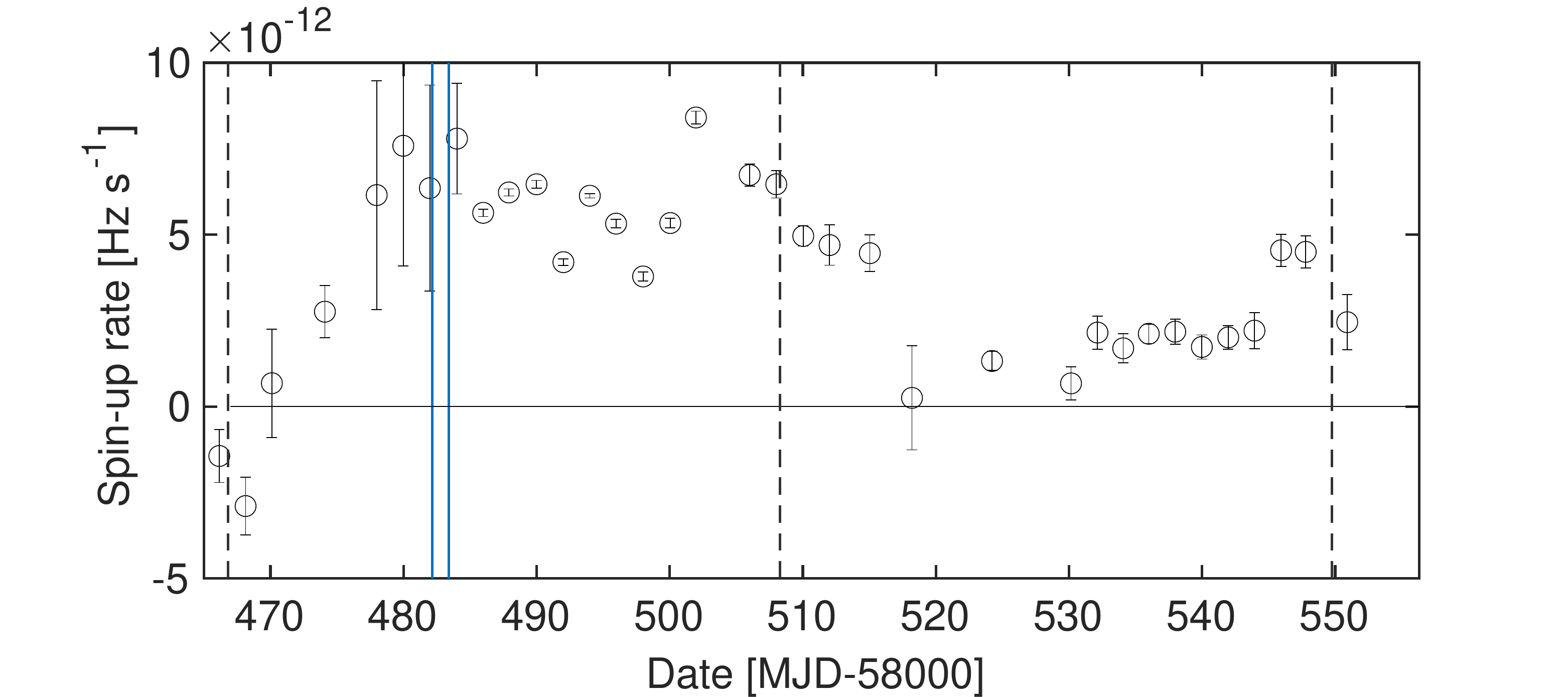}
\caption{The {\it Swift} BAT GX 301$-$2 hard X-ray
fluxes (upper panel), and 
GBM spin frequencies \citep[center, from the][]{Conn:19} and spin-up rates (bottom) 
for the two orbits with significant spin-up. The time period of the {\it X-Calibur} observations is shown by two vertical blue lines.  The periastron passages are marked by dashed vertical lines.
}
\label{f:spinup}
\end{figure}

The spin frequencies measured with the {\it Fermi} GBM in the \mbox{12-50\,keV} band (Fig. \ref{f:gbm}) 
show a spectacular spin-up coinciding with the exceptionally bright orbit.
During the orbit (41.5 days) covering the {\it X-Calibur} observations, the spin frequency (period) 
increased from 1.461\,mHz (spin period\,684\,s) on MJD 58,471.2 to 1.482\,mHz (spin period\,675\,s) 
on MJD 58,512.9 at a rate of 5.8$\times 10^{-12}$ Hz s$^{-1}$
\citep[see also][]{Nabi:19}.
The next orbit saw a much slower spin-up from 
1.482\,mHz on MJD 58,512.9 to
1.490\,mHz on MJD 58,553.2 at a rate of \mbox{2.3$\times 10^{-12}$ Hz s$^{-1}$}.
The spin-up rate is clearly correlated with an enhanced X-ray flux (Fig.\ \ref{f:spinup}), bolstering the
hypothesis that a change of the accretion 
rate or accretion mode is causing 
the spin-up.
Interesting features include the simultaneous dip
of the X-ray flux and spin-up rate at MJD 58,492, 
the decrease of the spin-up rate 
between MJD 58,502 and MJD 58,510
during a phase of rather constant
elevated X-ray flux levels, and 
the factor two lower spin-up during MJD 58,546 and MJD 58,548
when compared to the spin up one orbit earlier 
(MJD 58,503-MJD 58,510) at similar flux levels.

\citet{Koh:97} and \citet{Bild:97} reported similar spin-up phases detected with the BATSE gamma ray detectors. 
At the time, the spin frequency increased over 23 days (MJD 48,440-48,463) 
from 1.463\,mHz to 1.473\,mHz at a rate of 4.5$\times 10^{-12}$\,Hz\,s$^{-1}$
and over 15 days (MJD 49,245-49,230)
from 1.474\,mHz to 1.478\,mHz at a rate of 3.0$\times 10^{-12}$\,Hz\,s$^{-1}$.
All rapid spin-up periods were accompanied by heightened apastron activity.

\begin{figure}[t]
\epsscale{1.25}
\plotone{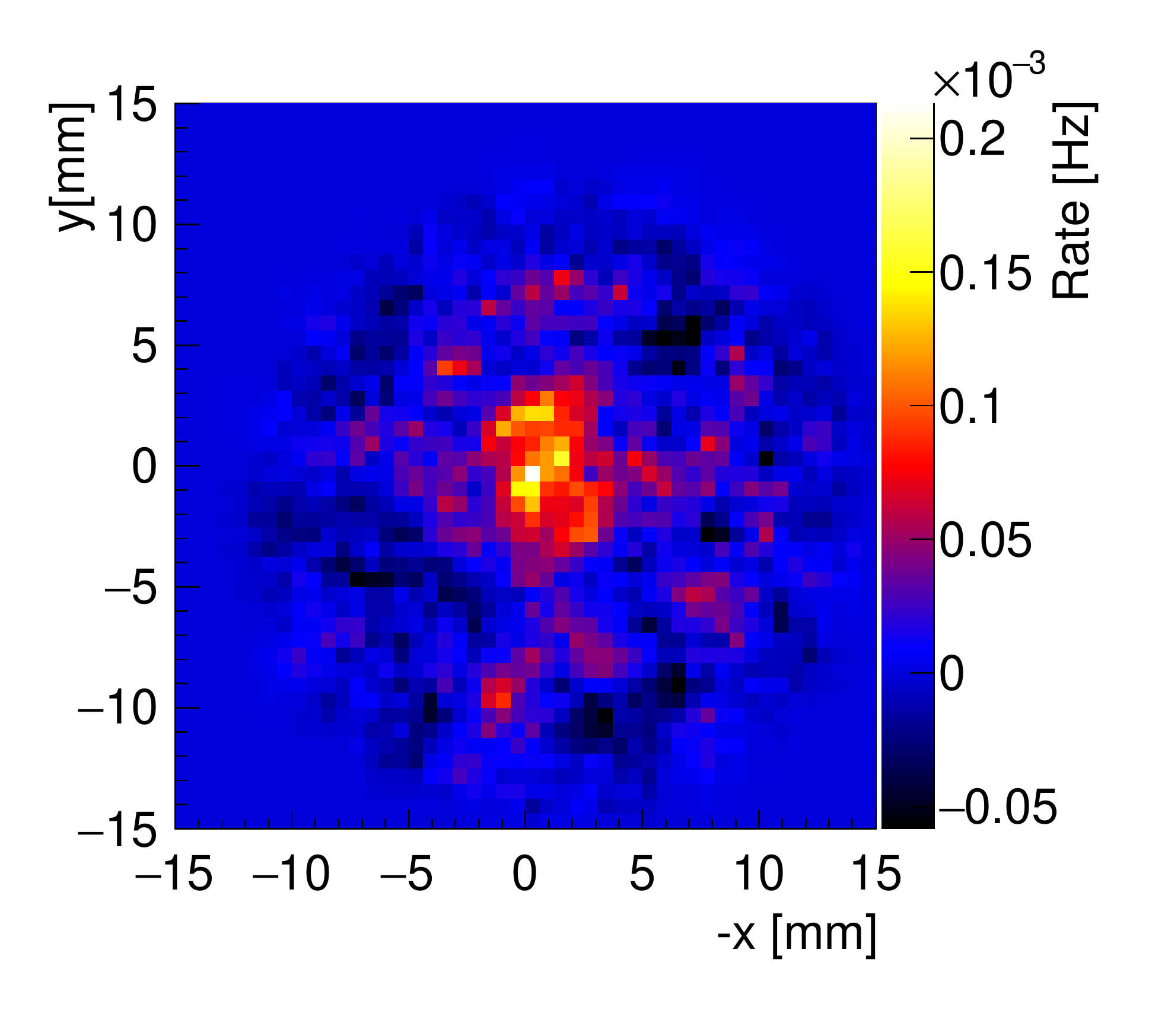}
\caption{\label{f:17} {\it X-Calibur} focal plane image of the X-ray pulsar GX 301$-$2 
recorded with the rear CZT detector (ON counts minus OFF counts). 
The image is referenced to the celestial North direction (up).
We only used half of the detector for this image, as the readout ASIC of the
second half worked only intermittently. This was the only ASIC (out of 34) showing 
problems during the flight. The spatial extent of the image is dominated 
by the mirror point spread function.
}
\end{figure}
\begin{figure}[t]
\epsscale{1.15}
\plotone{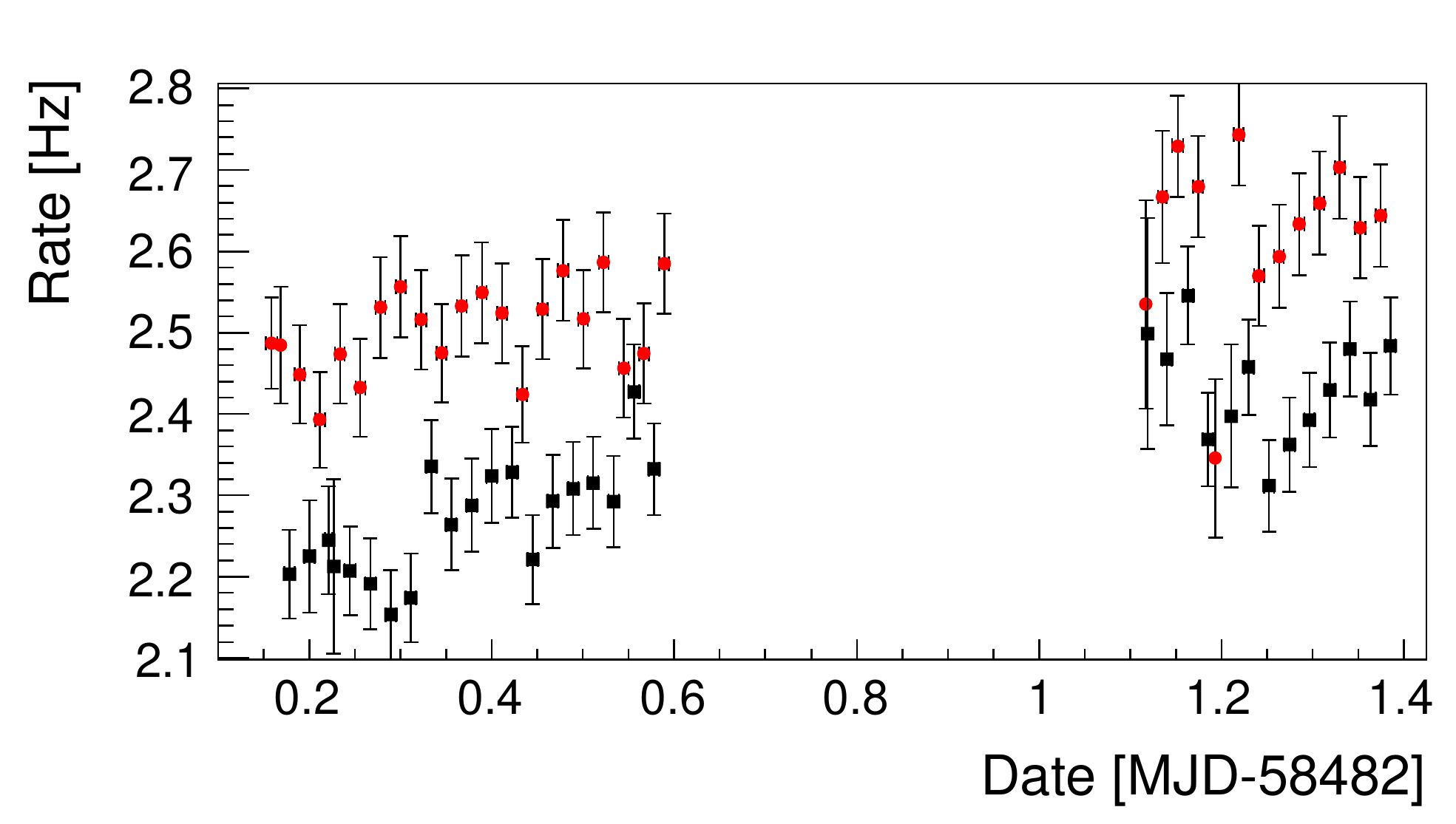}
\caption{\label{f:lc} {\it X-Calibur} 15-35\,keV detection rates on-source (red) and off-source (black) revealing an average source count rate of 0.23 Hz.
The rates are raw rates in the sense that they have not been corrected for the flight altitude and elevation-dependent 
atmospheric absorption.}
\end{figure}
\begin{figure}[t]
\epsscale{1.15}
\plotone{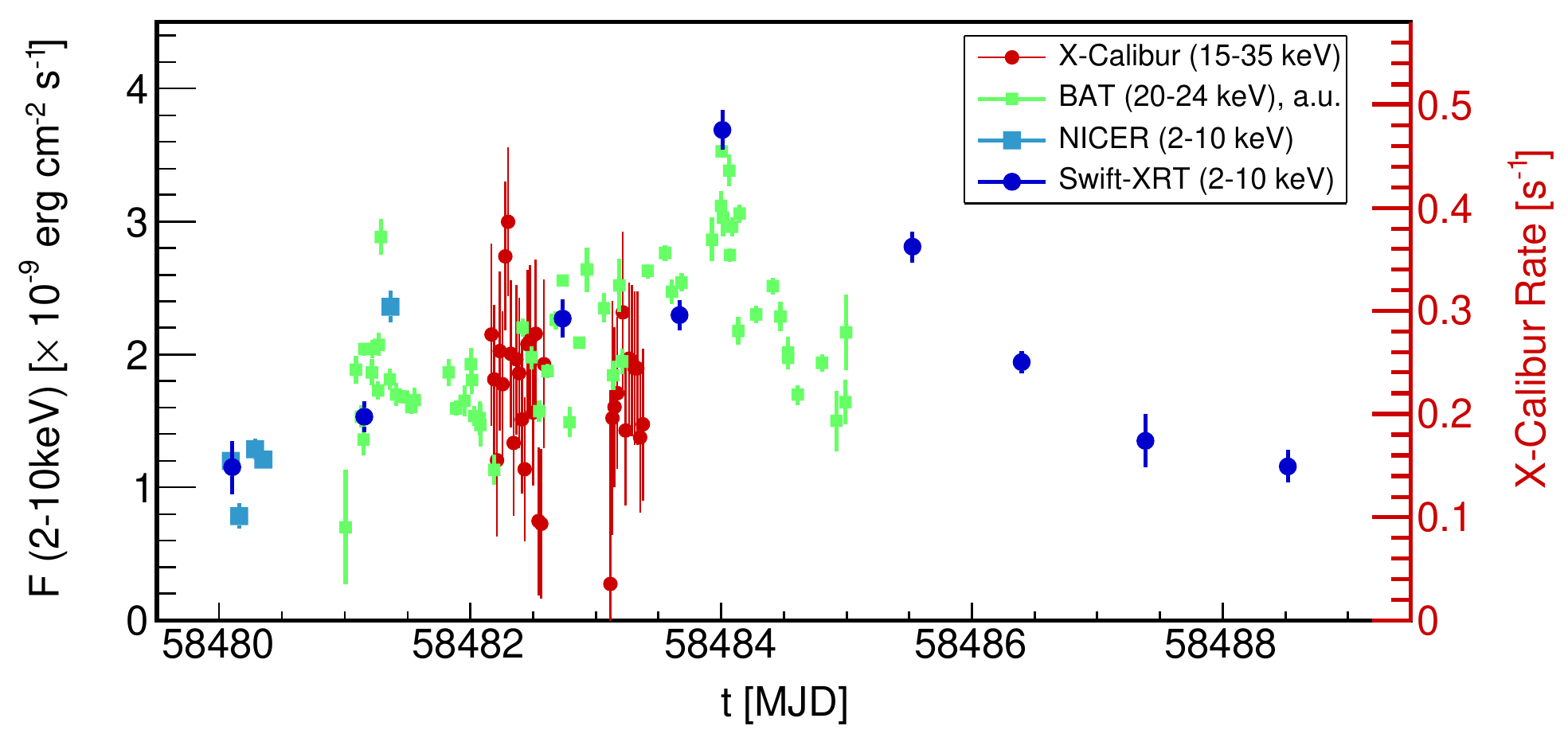}
\caption{\label{f:all} 
{\it X-Calibur} (15-35\,keV),
{\it Swift XRT} BAT (20-24\,keV),
{\it NICER} (2-10\,keV), and
{\it Swift XRT} (2-10\,keV) 
GX 301$-$2 detection rates around the 
apastron flare.} 
\end{figure}

Figure\,\ref{f:17} shows the GX 301$-$2 detection in the {\it X-Calibur} 
rear CZT detector. The image allows us to verify and refine the X-ray mirror alignment calibration (see also Appendix B).
Figure\,\ref{f:lc} presents the \mbox{15-35\,keV} ON and OFF light curves from the polarimeter section of the detector (without the rear CZT detector). Note that each data point corresponds to one \mbox{15-minute} run covering slightly more than one pulsar period. 
{\it X-Calibur} detected the source with a mean \mbox{15-35\,keV} rate of 0.23\,Hz.
Figure \ref{f:all} compares the X-ray light curves from {\it X-Calibur}, {\it Swift} BAT, {\it Swift} XRT and {\it NICER} 
taken around the time of the {\it X-Calibur} campaign. The flux level increased as the observation campaign unfolded and peaked a day after the {\it X-Calibur} observations ended. 
\begin{figure}[t]
\epsscale{1.15}
\plotone{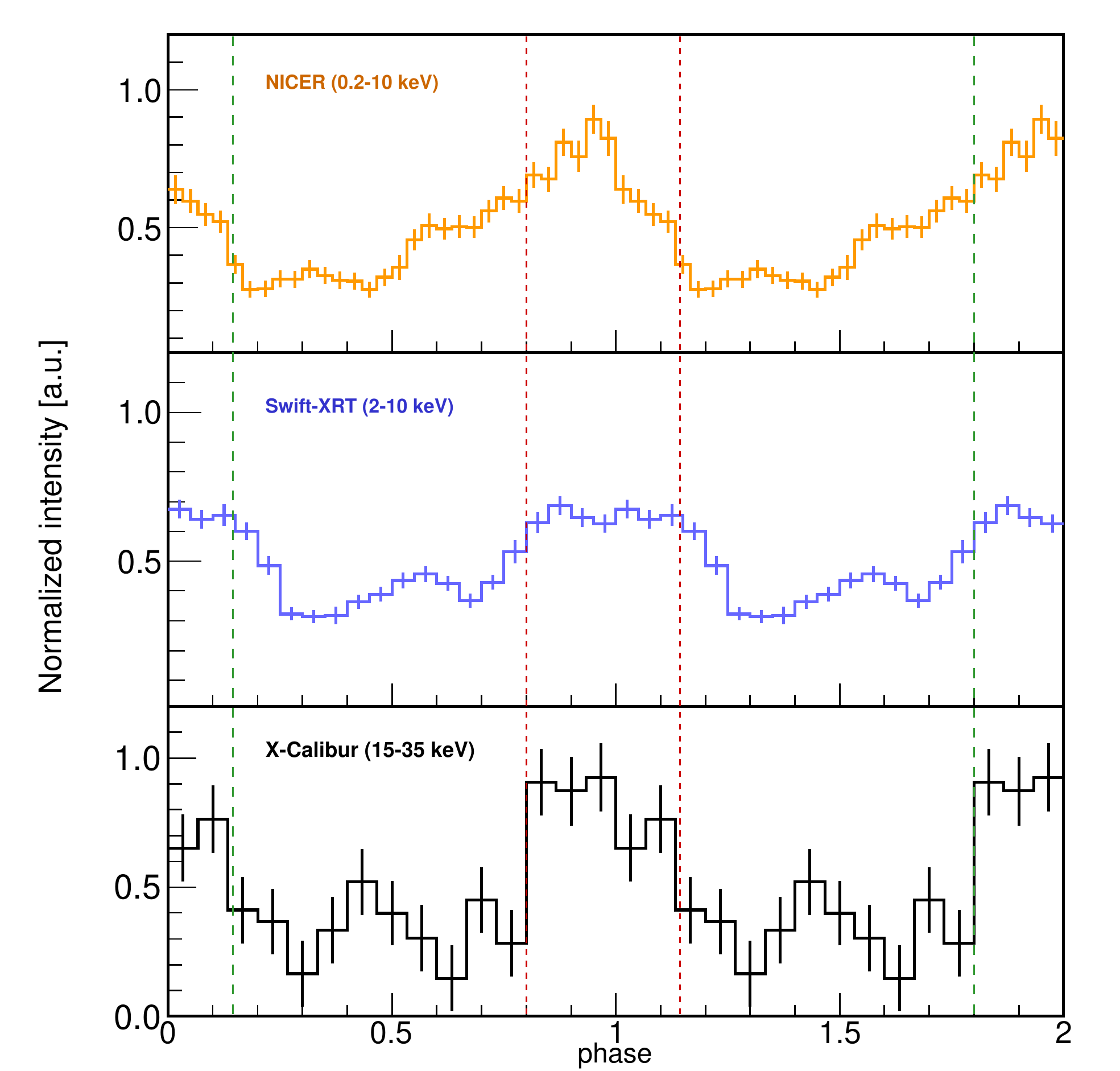}
\caption{\label{f:pulse} {\it NICER} (0.2-12\,keV), 
{\it Swift XRT} (0.2-10\,keV), and {\it X-Calibur} (15-35\,keV) 
time-averaged GX 301$-$2 pulse profiles. 
{ The vertical dashed line indicate the 
beginning and end of the main pulse 
(phase intervals 0.8-1.14).}}
\end{figure}

\begin{figure}[t]
\epsscale{1.15}
\plotone{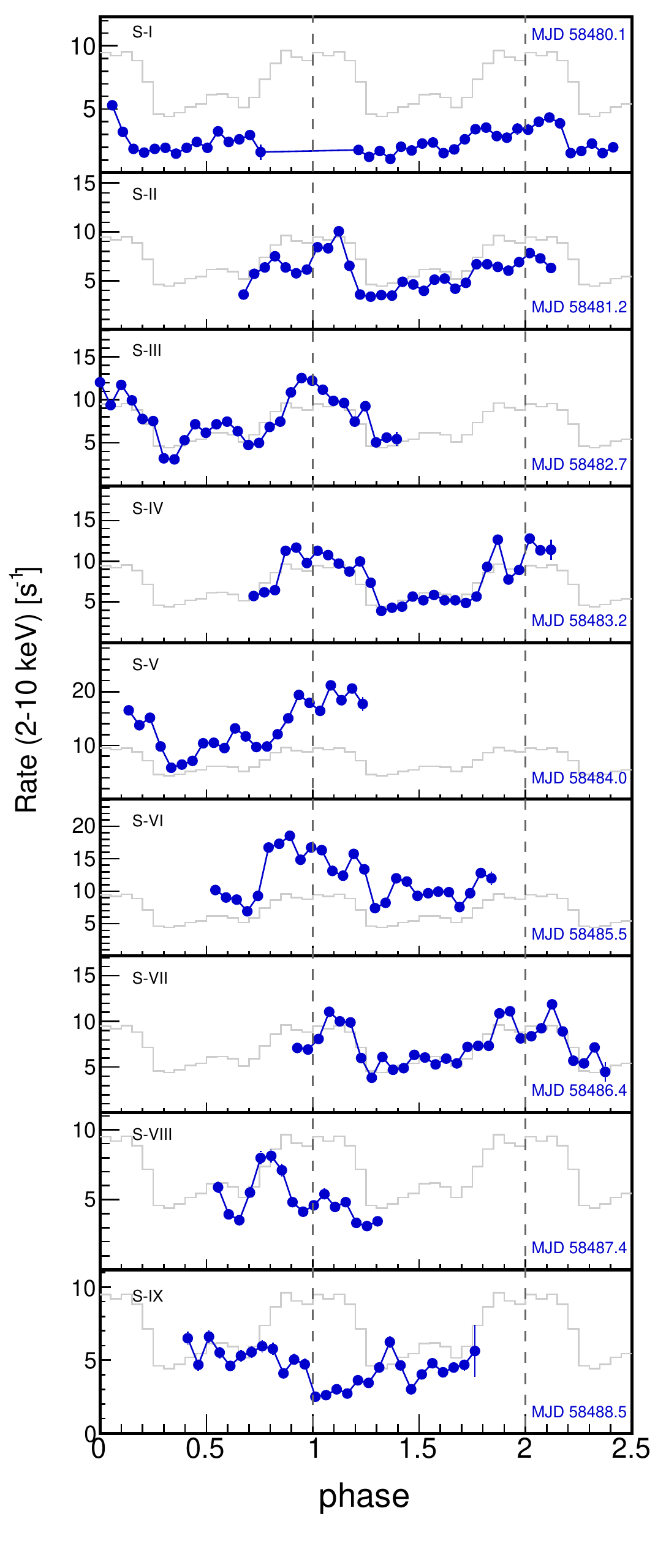}
\caption{\label{f:xrt}
Individual {\it Swift XRT} (0.2-10\,keV)
GX 301$-$2 pulse profiles showing large variations in pulse profiles 
(note the different scale on vertical axis). The light curve data is binned in 34 s lon time intervals. { The average XRT pulse profile is shown in grey for reference. The dashed vertical lines are shown to guide the eye.}
}
\end{figure}
Figure \ref{f:pulse} reports the average pulse profiles measured with the {\it Swift} XRT (0.2-10\,keV),
{\it NICER} (0.2-12\,keV), and  {\it X-Calibur} (15-35\,keV). 
All three pulse profiles show one peak strongly dominating over the other.
{ The shape of the {\it X-Calibur} 15-35\,keV pulse profile measured during the spin-up epoch deviates
significantly from the shapes of the 18-30\,keV pulse profiles recorded on 10/29/2014 (orbital phase 0.65),  10/4/2015 (orbital phase 0.85), and 3/3/2019 (orbital phase 0.89) 
with {\it NuSTAR}.  Whereas the {\it NuSTAR} 
pulse profiles show two pulses with approximately 
equal fluences 
\citep[flux integrated over time, see Figs. 3 and 4 of][]{Nabi:19}, the fluence of the main {\it X-Calibur} 
peak (phase 0.8-1.14) exceeds that of the secondary peak 1/2 period later by a factor of $\approx$2 with a 
statistical significance of more than 5 standard deviations.
The {\it NICER} (not shown here) 
and {\it Swift} data sets have sufficiently high signal-to-noise 
ratios to reveal significant variations of the pulse profiles from pulse to pulse
(Fig.\,\ref{f:xrt}). Such pulse profile variations can be caused by alterations in the accretion rate and by changes of the accretion and emission geometries.} 

{ 
\subsection{Spectral Results}}

The large absorption column observable in the {\it NICER} and {\it Swift} energy spectra reduces the count rate 
dramatically below 2\,keV. We select channels with energies between 2 and 10\,keV for spectral analysis, and fit them with a power law continuum going through a partially-covered absorber, and an additional Gaussian line:
\begin{equation} \label{e:2abs}
 \nh^\text{gal}  \times \biggl(\Bigl( c \nhone  + \left(1-c\right) \nhtwo \Bigr) \times \text{power law} + \text{Line}\biggr)
\end{equation}
where $\nh^\text{gal}$ was fixed to the galactic equivalent column density 
of $1.7\times10^{22}$\,cm$^{-2}$ reported in \citep{kalberla05a}. 
\begin{figure}[tbh]
\epsscale{1}
\includegraphics[angle=-90,width=3in]{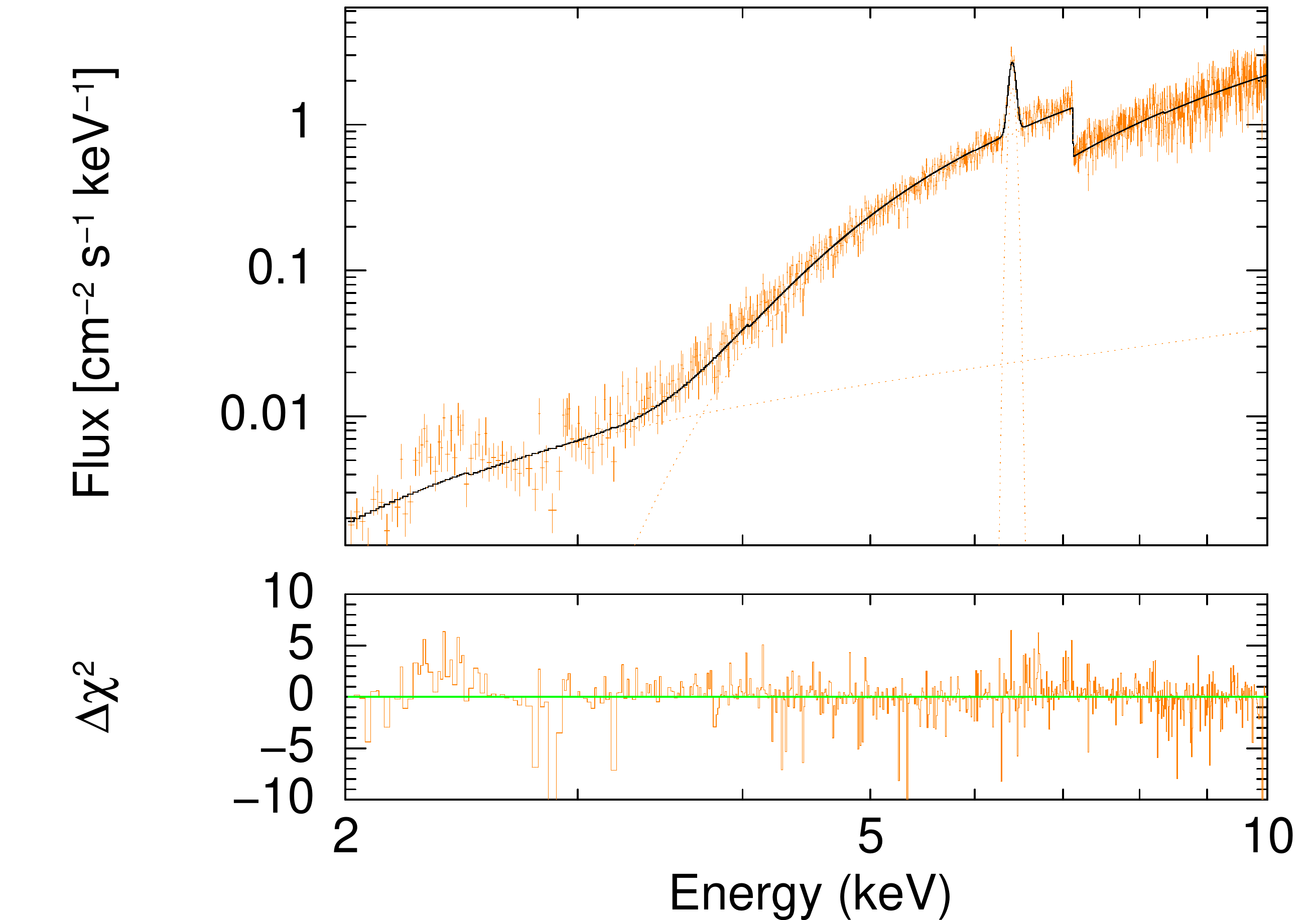}
\includegraphics[angle=-90,width=3in]{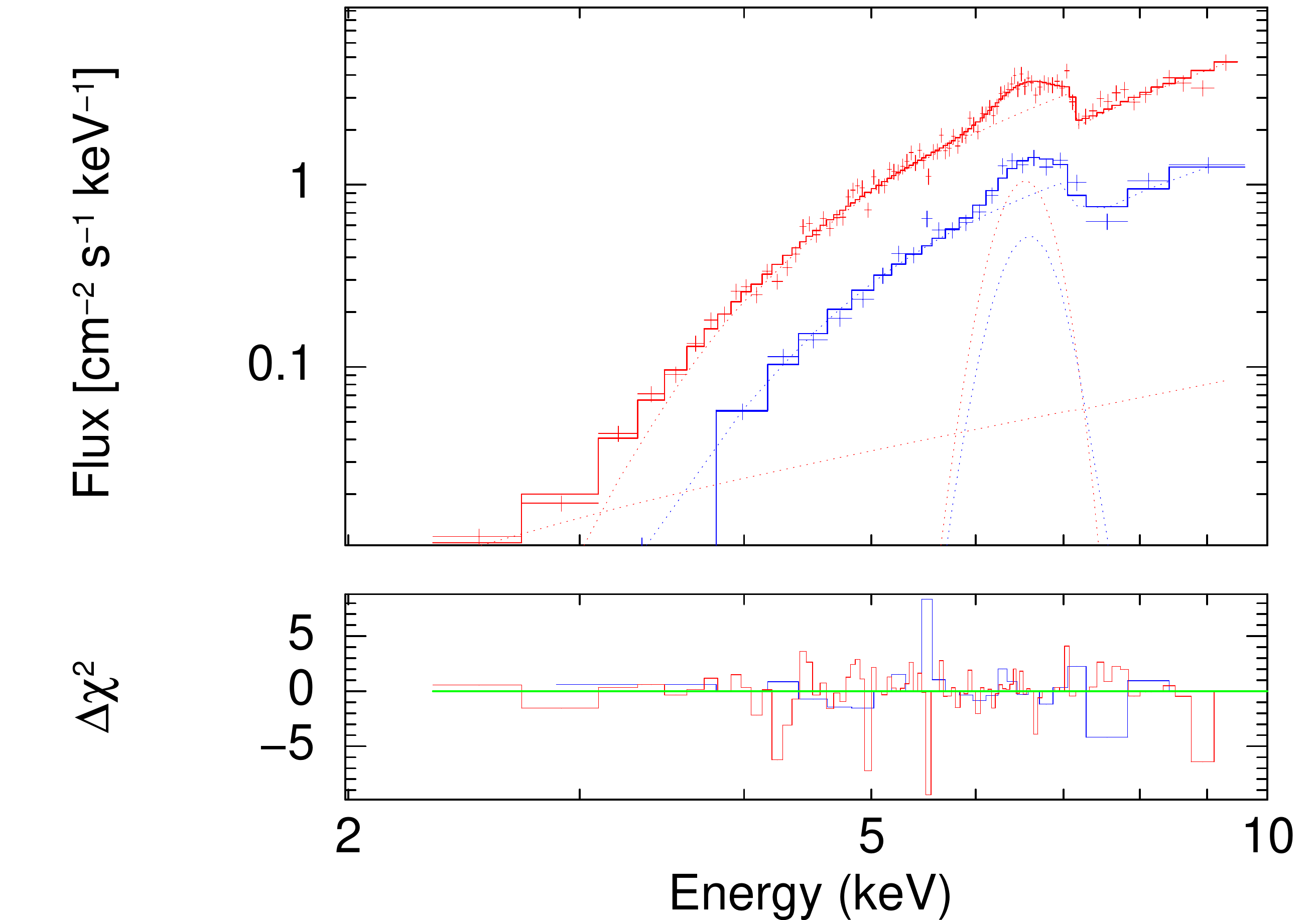}
\includegraphics[angle=-90,width=2.9in]{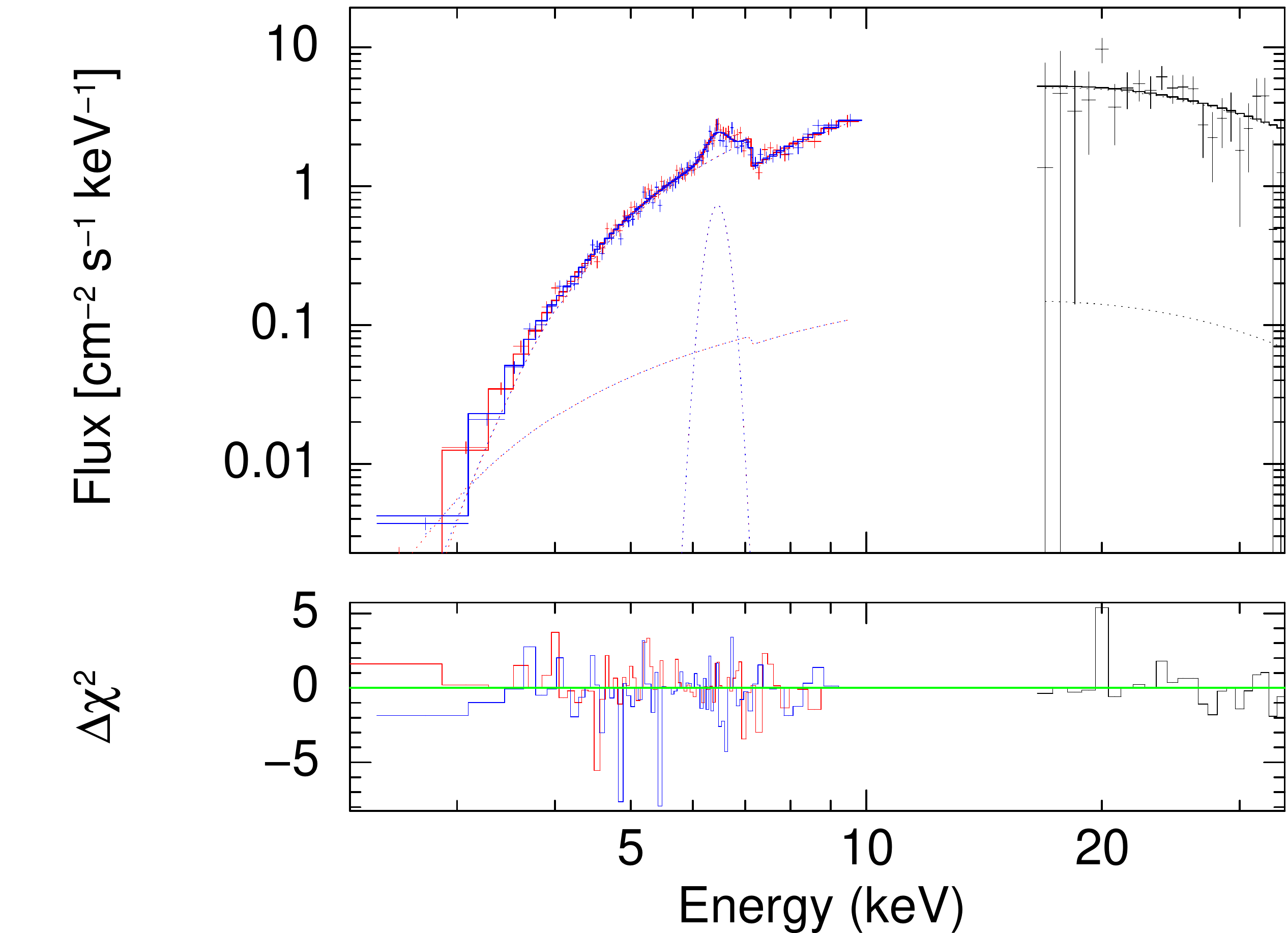}
\caption{\label{f:nicerSpec} 
Top: {\it NICER} GX 301$-$2 energy spectrum measured on MJD 58,480.34 
(observation {\it N-IV} from Table \ref{tabO}). 
Middle: {\it Swift}-XRT spectra from observation {\it S I} (in blue) and {\it S V} in red (see Tables  \ref{tabO} and \ref{tabN} for details).  
Bottom: Joint {\it Swift XRT} (observations {\it S III} and {\it S IV} from Table \ref{tabO}) and {\it X-Calibur} GX 301$-$2 energy spectrum 
measured on MJD 58,482 and MJD 58,483. The two top panels also show the best-fit model
(solid lines) and the model components (dashed lines) from Equ.\,(\ref{e:2abs}), while the bottom panel displays the mode components from Equ.\,\ref{e:exp}. }
\end{figure}

The results are reported in Tables \ref{tabN}-\ref{tabS}. 
Given the wide variation of the  signal-to-noise ratios of the different
data sets, some of the energy spectra 
do not constrain some of the parameters of the model from Equ.\,\ref{e:2abs}. In those cases, the parameters without errors in Tables \ref{tabN}-\ref{tabS} were fixed to the reported values during the fitting process. For example, in observation $S I$ (Figure\,\ref{f:nicerSpec}, top, and Table\,\ref{tabS}) the data do not allow us
to constrain the second absorption component, so we fit the spectrum using Equ.\,\ref{e:2abs} with $c=1.0$ and $\nhtwo=0$.
For the main absorbing component we find  $N_{\rm H}$ values of between 
\mbox{$\sim$33$\times 10^{22}$ cm$^2$} and 
\mbox{$\sim$90$\times 10^{22}$ cm$^2$}.
The $N_{\rm H}$ of the main component decreases through the apastron flare 
until MJD 58,485.52 (observation {\it S VI}). Most of our 
values are higher than the pre-periastron column densities of between   
\mbox{$\sim$15$\times 10^{22}$ cm$^2$} and 
\mbox{$\sim$40$\times 10^{22}$ cm$^2$} from \citet{Such:12,Fuer:18},
and lower than the periastron values of between  
\mbox{$\sim$115$\times 10^{22}$ cm$^2$} and 
\mbox{$\sim$175$\times 10^{22}$ cm$^2$} of \citet{Fuer:11}.

The {\it NICER} energy spectra show a clear
Fe K$\alpha$ lines, and some 
marginally significant deviations of the data from the best-fit model between 
2\,keV and 3\,keV (Figure\,\ref{f:nicerSpec}, top). The {\it Swift}-XRT spectra also show the presence of the Fe K$\alpha$ line (Figure\,\ref{f:nicerSpec}, middle) throughout the whole observation period.

The {\it X-Calibur} 15-35\,keV energy spectrum is fitted with a power law
model. We obtain a \mbox{15-35\,keV} flux of \mbox{$(7.4^{+1.4}_{-1.3}) \times 10^{-9}$ erg cm$^{-2}$ s$^{-1}$}
and a power law index of \mbox{4.2 $\pm $ 0.6} (1$\sigma$ errors).
{ The photon index agrees within statistical errors with the energy spectrum measured 
with {\it NuSTAR} on 3/3/2019 which exhibits 
a rollover from a photon index 
of $\Gamma=2$ at 20 keV to $\Gamma=4$ at 30 keV
\citep[Fig.\,6 of][]{Nabi:19}.}

We study the broadband $2-35$\,keV energy spectrum by simultaneously fitting 
the {\it Swift}-XRT (observations
{\it S III} and {\it S IV}) and {\it X-Calibur} data (Figure\,\ref{f:nicerSpec}, bottom) with a power-law model with an exponential cutoff, 
a partially-covered absorber, 
and an additional Fe-K$\alpha$ fluorescence line:
\begin{eqnarray}
\label{e:exp}
     \nh^\text{gal} \times \biggl(\Bigl( c \nhone  + \left(1-c\right) \nhtwo \Bigr) \times \nonumber  \\
     \times E^{-\Gamma}\exp{\left( -E/E_{\rm fold} \right)} + \text{Line}\biggr).
\end{eqnarray}
A model with $\Gamma=0.04 \pm 0.21$, $E_{\rm fold} = 7.95\pm0.78$\,keV and $N_{\rm H,1}=(56\pm8)\times 10^{22}$\,cm$^2$ gives a good fit to the broadband data, with $\chi^{2}/{\rm NDF} = 159.2/155$. The values of the spectral parameters are similar to those obtained by \citet{Fuer:18} using {\it NuSTAR} observations, with the exception of the softer photon index of $\Gamma\sim0.8$ obtained by \citet{Fuer:18}.\\
\subsection{{\it X-Calibur} Polarization Analysis}
\begin{figure}[t]
\epsscale{1.2}
\plotone{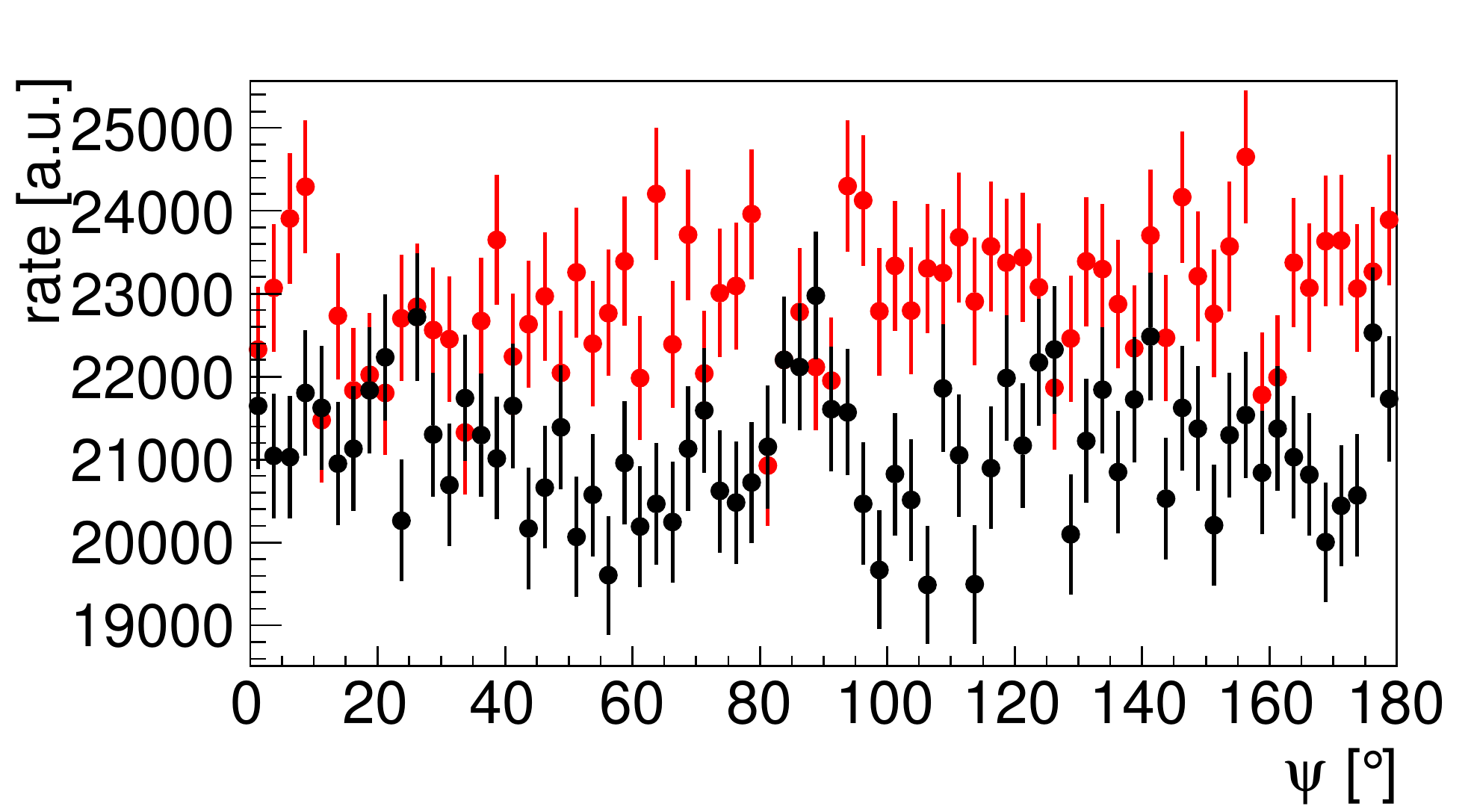}\vspace*{-0.1cm}
\plotone{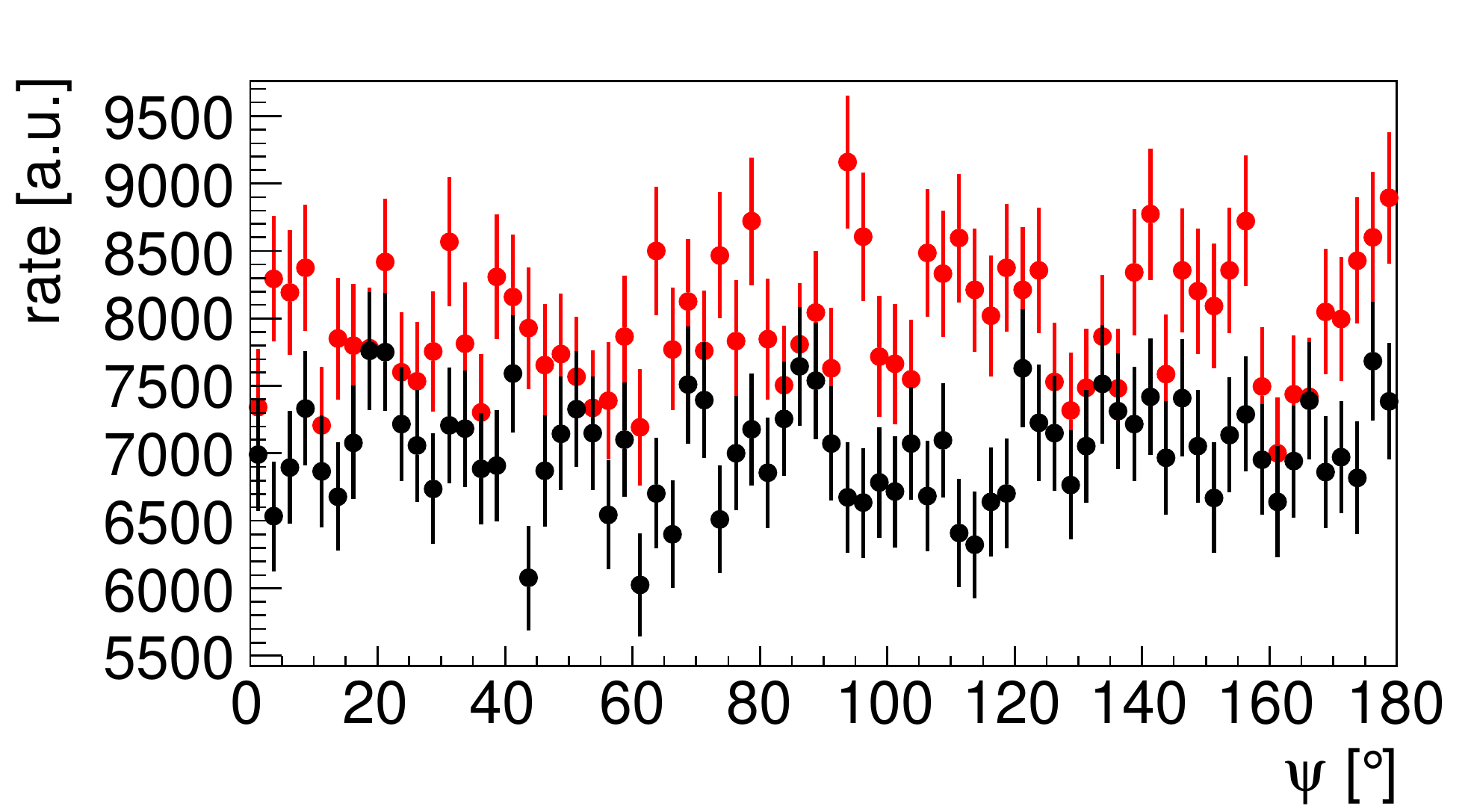}\vspace*{-0.1cm}
\plotone{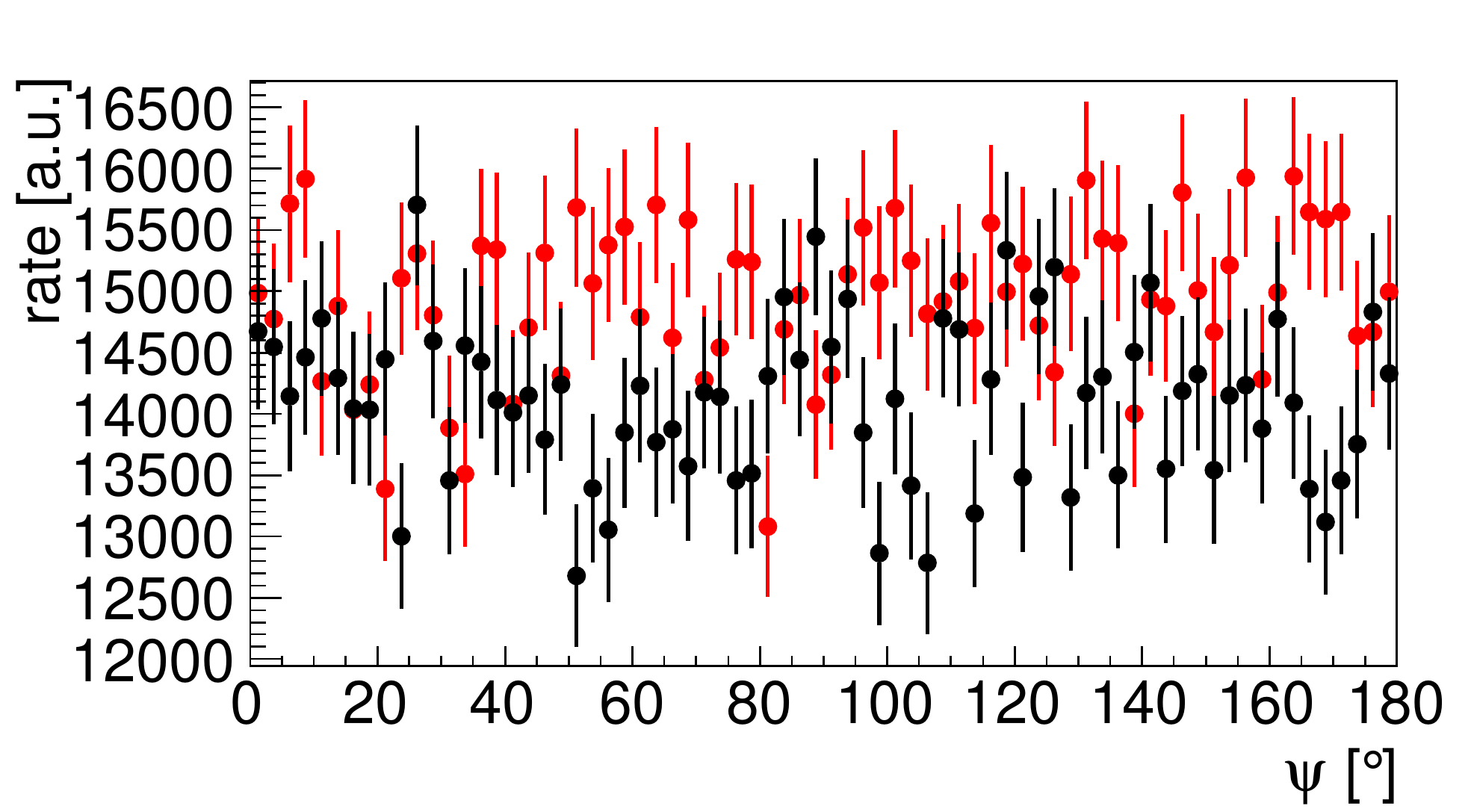}\vspace*{-0.1cm}
\caption{\label{f:chiOnOff} Distribution of the azimuthal scattering angles for the entire emission (top panel),
the main pulse (phase 0.8-1.14, center panel), 
and for the bridge and secondary pulse emission 
(phase 0.14-0.8, bottom panel) 
for the ON (red) and OFF (black) data.
Individual events enter the analysis with a weight, and we thus give the rate per bin (i.e.\ the weighted number of events per unit time per bin) in arbitrary units (a.u.).}
\end{figure}
All polarization results are given in the 15-35\,keV band for three data sets 
(see lower panel in Fig.\,\ref{f:pulse}): 
(i) the entire data set, 
(ii) the main pulse (pulsar phase 0.8-1.14), and for (iii) the bridge and secondary pulse emission (pulsar phase 0.14-0.8).
Figure \ref{f:chiOnOff} presents the modulation curves (azimuthal scattering angle distributions) for the ON and OFF observations. Neither the ON nor the OFF distributions show obvious
modulations. 

\begin{figure}[t]
\epsscale{1.25}
\plotone{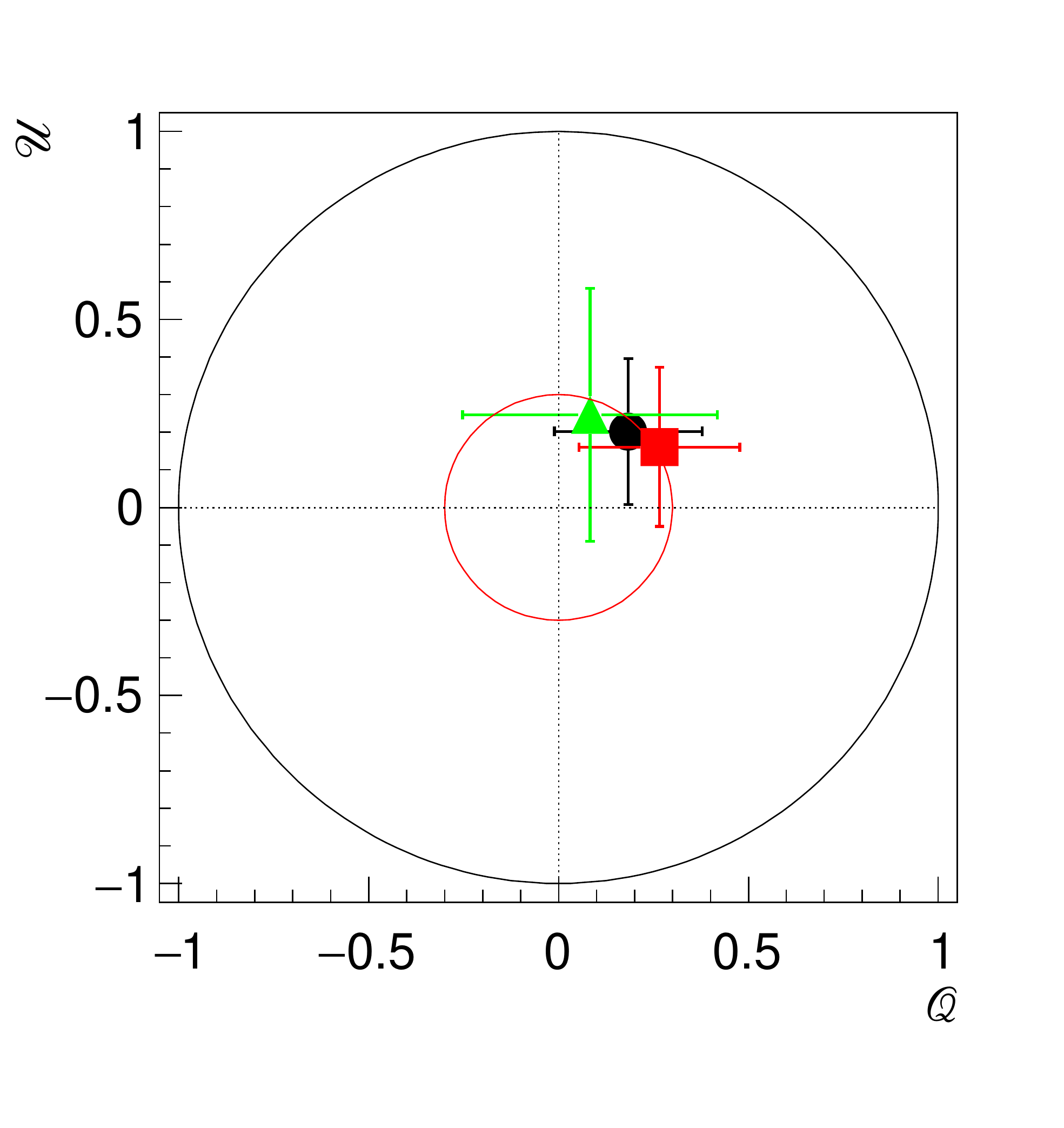}
\vspace*{-1cm}
\caption{\label{f:xpol1} {\it X-Calibur} constraints on the 
linear polarization of the 15-35\,keV GX 301$-$2 emission in the plane of the normalized Stokes parameters for the entire data set (black filled circle), the main pulse (red square, phase 0.8-1.14), and the bridge and secondary pulse (green triangle, phase 0.14-0.8) with 1$\sigma$ statistical errors.
Polarization fractions of 0\%, 30\% (for illustrative purposes), 
and 100\% correspond to ${\cal Q}={\cal U}=0$ point at the center of the graph, 
the red circle, and the black circle, respectively.}
\end{figure}

Figure\,\ref{f:xpol1} presents the results in 
the ${\cal Q}$-${\cal U}$ plane for all three data sets. The statistical significance for a polarization detection can be calculated 
with $Q$ and $U$ which
have slightly smaller relative errors than
${\cal Q}$ and ${\cal U}$.
The overall results deviate by $\sqrt{(Q/\sigma_Q)^2+(U/\sigma_U)^2}\,=$
1.41 (entire emission), 1.47 (main pulse), and 0.78 (bridge and secondary pulse) standard deviations from 
zero polarization ($Q=0$ and $U=0$). 
The {\it X-Calibur} observations thus did 
not lead to a significant detection of a 
non-zero polarization. 

\begin{figure}[t]
\epsscale{1.25}
\plotone{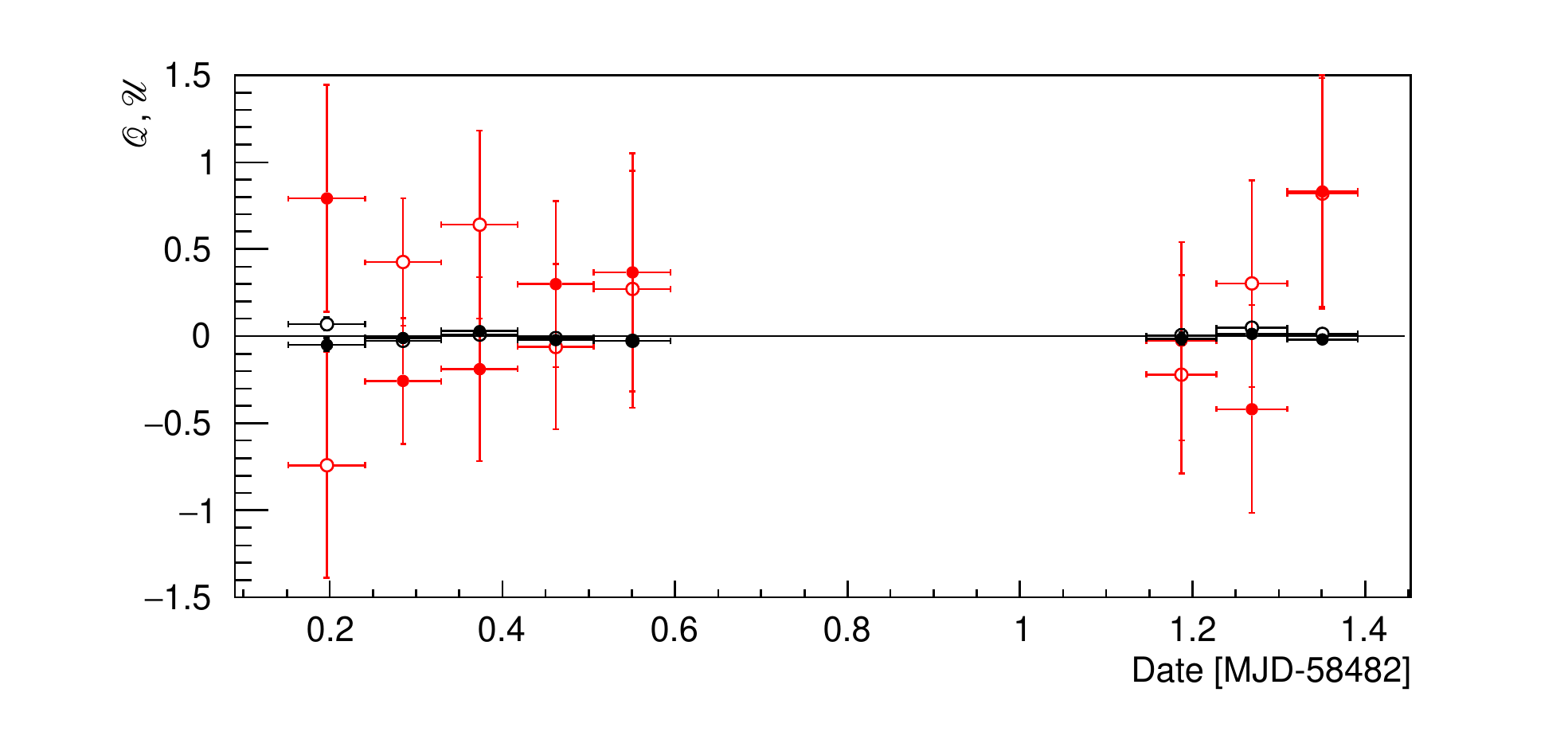}\vspace*{-0.5cm}
\caption{\label{f:qu}
${\cal Q}$ (filled circles) and ${\cal U}$ (open circles) parameters 
for the background-subtracted ON-data (red) and the OFF-data (black)
as a function of time.
}
\end{figure}

For the pulse-integrated emission, Fig.\,\ref{f:qu} shows the ${\cal Q}$ 
and ${\cal U}$ parameters for the background-subtracted ON-data 
and the OFF background data as a function of time. It can be seen that the 
${\cal Q}$ and ${\cal U}$ parameters 
of the ON and OFF observations 
are consistent with zero 
polarization for all time intervals.
The same applies to the
${\cal Q}$ and ${\cal U}$ parameters 
of the entire OFF data set.

\begin{figure*}[t]
\plotone{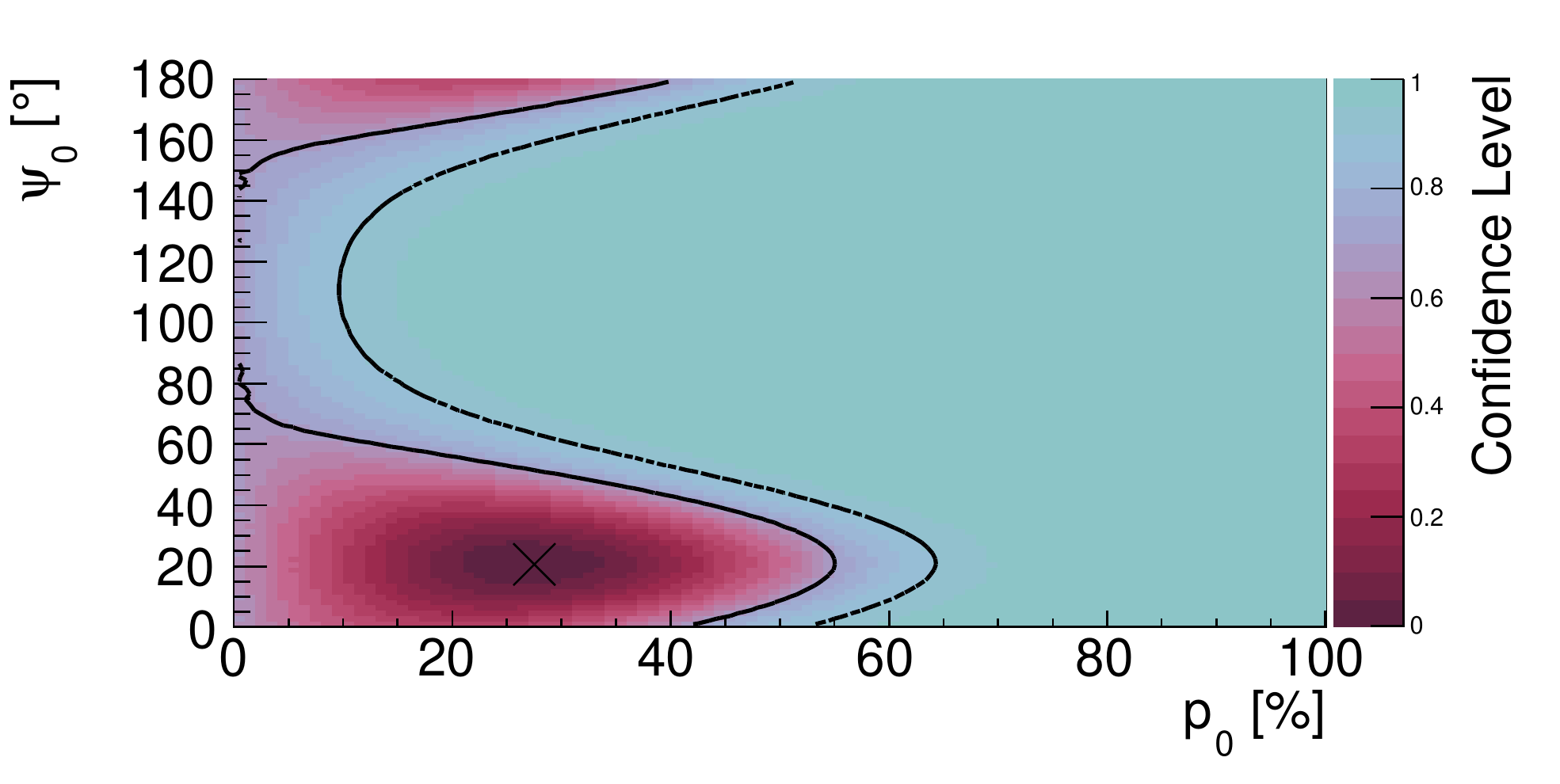}\vspace*{-1.0cm}
\plotone{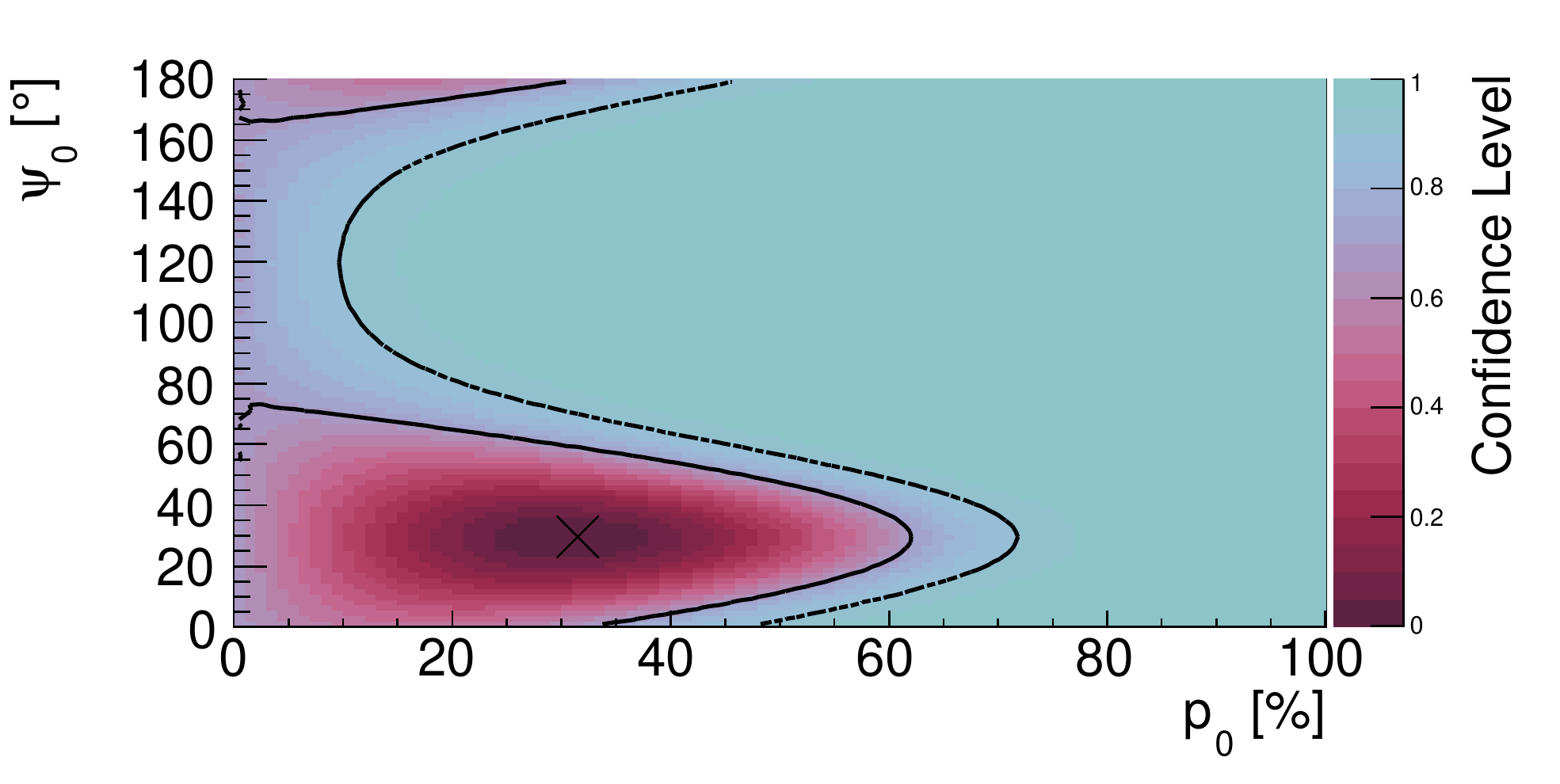}\vspace*{-1.0cm}
\plotone{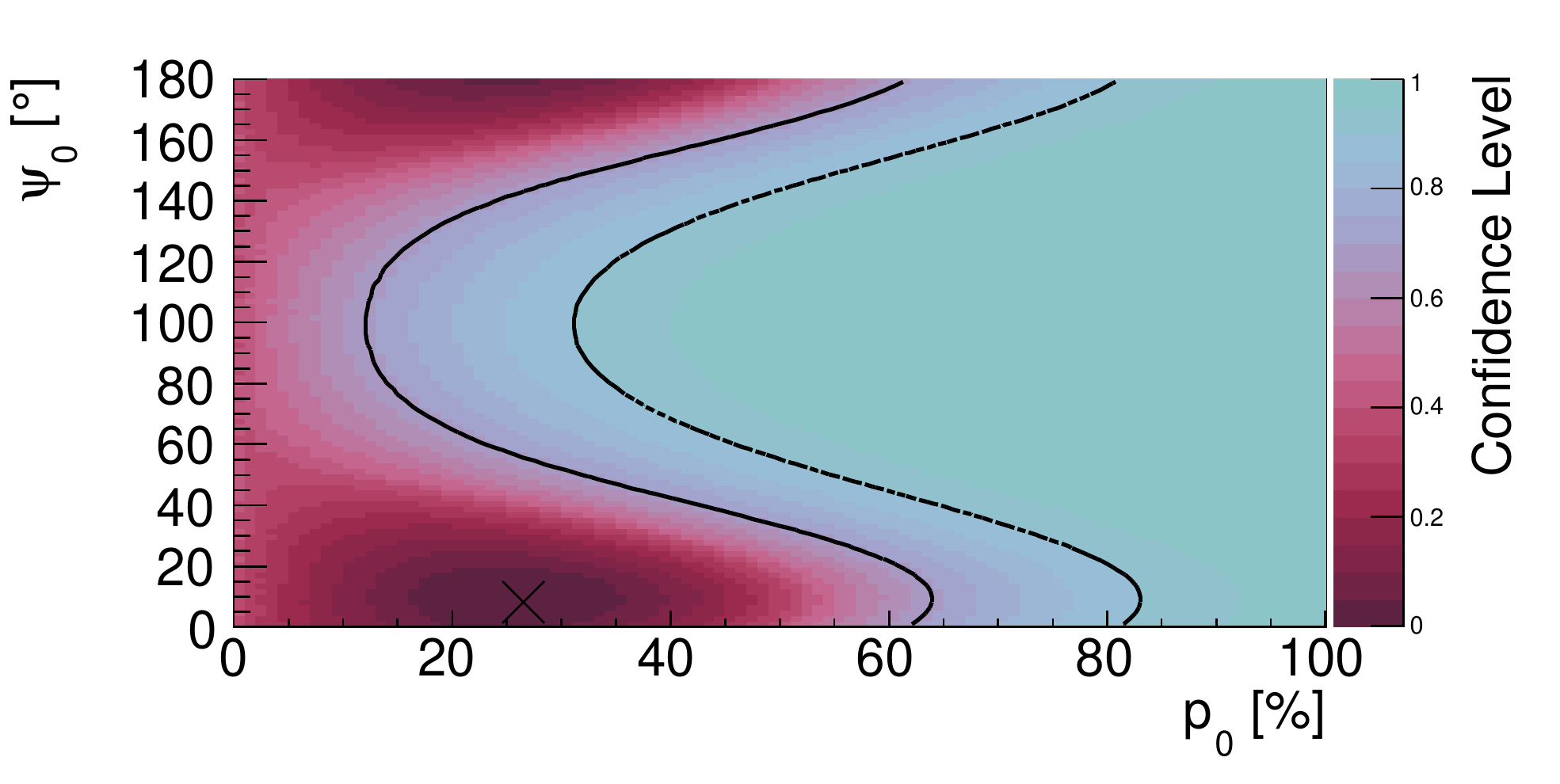}\vspace*{-0.2cm}
\caption{\label{f:xpol2}
{\it X-Calibur} 15-35\,keV polarization results in the 
polarization fraction $p_0$ and 
polarization angle $\psi_0$ plane
for the entire emission (top),
the main pulse 
(pulsar phase 0.8-1.14, center), and the
bridge emission and the secondary pulse 
(pulsar phase 0.14-0.8, bottom).
The most likely $p_0$-$\psi_0$ combination 
is marked by a cross.
The color scale shows the results for different confidence levels, and the contours delineate 
the 68.27\% (1 $\sigma$) 
and 90\% confidence regions. 
The analysis only accounts for statistical errors.}
\end{figure*}

\begin{deluxetable*}{ccccccc}[tbh]
\tabletypesize{\scriptsize}
\tablecaption{{\it X-Calibur} 15-35\,keV polarization results. Errors on $p$ and $\psi$ are on 90\% Confidence Level.
The polarization angle of the third data set is unconstrained on the 90\% confidence level.\label{t:pol-res}}
\tablewidth{0.pt}
\tablehead{
\colhead{Phase Interval} & \colhead{$\cal Q\,[\%]$}& \colhead{$\cal U\,[\%]$}&  \colhead{Deviation from $p=0$ [$\sigma$]} & 
\colhead{$p$ [\%]} & 
\colhead{$\psi$ [$^{\circ}$]} & 
\colhead{Upp.\ Lim.\ $p$ (90\% CL)} [\%]}
\startdata
All (0-1) & 18.4 $\pm$ 19.4 & 20.2 $\pm$ 19.4 & 1.41 & 27$^{+38}_{-27}$ & 21 $\pm$ 43 & 46.9 \\
Main Pulse (0.8-1.14) & 26.6 $\pm$ 21.2 & 16.1 $\pm$ 21.1 & 1.47 & 32$^{+41}_{-32}$ & 30 $\pm$ 40 & 52.3\\
Bridge and Sec. Pulse (0.14-0.8) & \phantom{1}8.3 $\pm$ 33.5 & 24.6 $\pm$ 33.6 & 0.78 & 27$^{+55}_{-27}$ & 10 & 62.2 
\enddata
\end{deluxetable*}
Figure\,\ref{f:xpol2} presents the observational constraints on the polarization fraction $p_0$ and angle $\psi_0$.
We use a Bayesian analysis with a 
flat prior of the polarization fraction $p_0$ 
between 0\% and 100\% 
and the polarization angle $\psi_0$ 
between 0 and $\pi$ \citep{Quin:12,Kisl:15}:
\begin{eqnarray}
dP_0(p_0,\psi_0) & = & \, \text{const}\,dp_0\, d\psi_0 \, \nonumber\\[-5.5pt]
\label{eq:Bayesian_prior}\\[-5.5pt]
 & \propto & 1/\sqrt{{\cal Q}^2+{\cal U}^2}\, d{\cal Q}\,d{\cal U} \quad .
 \nonumber
\end{eqnarray}
The most likely true parameter combination $p_0$ and $\psi_0$ is shown by a cross mark, and the confidence regions are shown by contours and the color scales. Table \ref{t:pol-res} lists the most likely values of $p_0$ and $\psi_0$
together with the confidence intervals derived from the distributions in Fig.\ \ref{f:xpol2}. The table includes 
the 90\% confidence interval upper limits 
on the polarization fraction $p_0$ calculated by marginalizing 
the probability density function $P(p_0,\psi_0)$ over $\psi_0$. 


\section{Summary and Outlook}  
 \label{sec:discussion}
This paper presents the results of the observations of the accretion-powered X-ray pulsar GX 301$-$2 with {\it X-Calibur}, 
{\it NICER}, the {\it Swift} XRT, and BAT, and {\it Fermi} GBM.
The observations reveal a rare flaring period between the periastron flares associated with
a spin-up of the pulsar similar to earlier events 
\citep{Koh:97,Bild:97}. Historically, the spin of GX 301$-$2 exhibited values around 
1.4 mHz (pulsar period: 715\,s) between 1975 and 1985 and values around the current value of 1.47 mHz
(pulsar period: 680\,s) between 1993 and now
\citep{Whit:76,Naga:89,Luto:94,Koh:97,Bild:97}, indicating an approximate equilibrium between spin-up and spin-down torques 
during these two long epochs \citep{Lipu:92,Doro:10}.  
The spin-up epochs typically last about one orbit, in which the pulsar frequency changes linearly.
The spin-up period starting at the time of the {\it X-Calibur} observations lasted for two orbits with a marked decline of the spin-up rate after the first orbit.   
The spin-up events start briefly after periastron 
\citep[Fig.\,\ref{f:gbm} in this paper, and Fig.\,11 of][]{Koh:97}.  

A possible interpretation of these signatures is that the neutron star acquires a temporary accretion disk \citep{Koh:97} 
shortly after periastron passage. The temporary disk provides 
fuel for one orbit during which the 
pulsar spins up continuously, and is destroyed during the next periastron passage. 
The disk may form for example when the neutron star crosses the plasma stream from { Wray 977} at the orbital 
phase of $\sim$0.25 \citep{Leah:08}.
We note that during the spin-up periods, the crossing always results 
in a large increase in X-ray flux at the orbital phase of $\approx$0.4.
Independent of what exactly triggers the X-ray flares, 
it is an open question why only some flares spin up the neutron star.

We report here on the first constraints on the hard X-ray polarization of an 
accreting neutron star at energies fairly close to the cyclotron line energy.  
Owing to the short balloon flight time, the {\it X-Calibur} observations 
did not yield a definitive polarization detection, but did offer constraints 
on the polarization fraction and the polarization angle plane. 
The results can be compared to the 
predictions from \cite{Mesz:88}.
The authors find that the propagation 
of the radiation 
in the ordinary and extraordinary mode and the 
strongly mode-dependent scattering cross-sections 
can lead to very high ($\sim$80\%) polarization 
fractions for certain pulse phases 
close to the cyclotron resonant energy.  
Interestingly, they find that fan-beam models predict 
rather robustly a positive correlation of the peak 
intensity and the polarization fraction. 
In contrast, pencil-beam models predict the opposite: 
a minimum (maximum) of the polarization fraction during the peak (valley) of the pulsed emission. 
\begin{figure}[t]
\epsscale{1.2}
\plotone{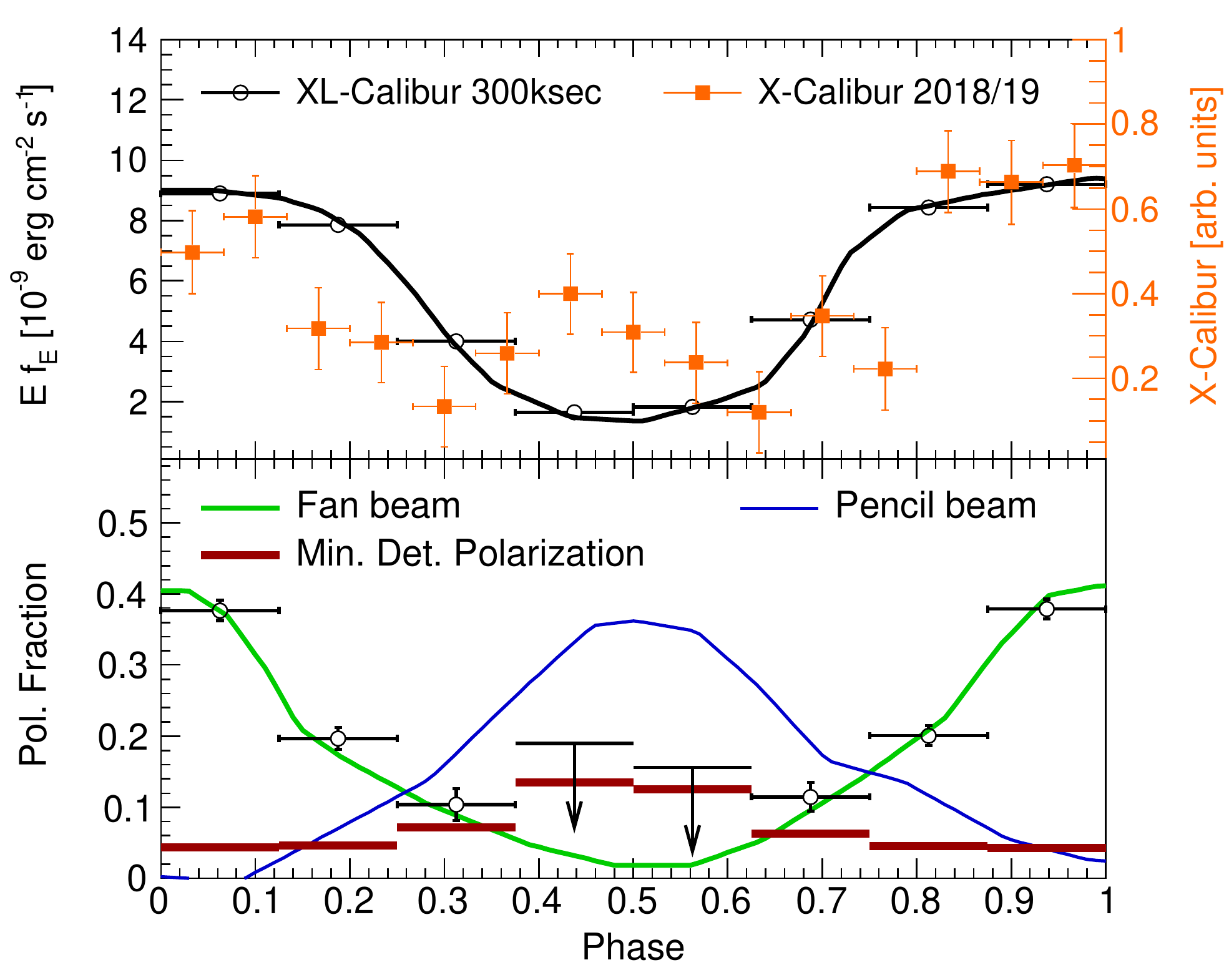}
\caption{\label{f:xlc} Simulated outcome of a 300 ksec GX 301$-$2 observation with {\it XL-Calibur}, assuming a 20-50\,keV flux of 700\,mCrab, an energy spectrum similar to those from \citet{Fuer:18}, and an atmospheric 
depth of 7 g/cm$^2$ (equal to the 
mean depth of the 2018/2019 GX 301$-$2 observations).
Top: assumed pulse profile (black line), measured {\it X-Calibur} 2018/19 pulse profile (orange data points), and simulated {\it XL-Calibur} results (black data points).   
Bottom: Expected polarization fractions  for the fan beam (green line) and pencil beam (blue line) models of \citet{Mesz:88} (model “45/45”). 
The black data points show the simulated {\it XL-Calibur} polarization fraction results for the fan beam model, and the dark red lines show the Minimum Detectable Polarizations (MDPs), i.e.\ the polarization fractions 
that {\it XL-Calibur} could detect with a 99\% confidence level. }
\end{figure}
The {\it X-Calibur} observations constrain the polarization fraction in the 15-35\,keV band, 
somewhat below the centroids of the CRSFs at 35\,keV and 50\,keV. The calculations of \citet{Mesz:88} were carried out for a cyclotron resonance at 35\,keV. 
At 25\,keV the pencil-beam (fan-beam) model predicts polarization fractions of $\sim$20\% ($<$5\%).  The {\it X-Calibur} GX\,301$-$2 result of $p_{\rm r} = 27^{+38}_{-27}\,$\% cannot distinguish between the two models.  Doing so with high statistical certainty will require future observations with a one-sigma error of \mbox{$<$4\%}.  

Driven by this requirement, we are now working on an {\it X-Calibur} follow-up mission called {\it XL-Calibur} \citep{Abar:19b} which promises hard X-ray polarimetric 
observations with one to two orders of magnitude improved signal-to-background ratio.
The mission uses the 12\,m focal length mirror 
fabricated for the Formation Flight Astronomical 
Survey Telescope (FFAST) \citep{Tsun:14} 
which offers more than three times larger effective areas than
the current mirror \citep{Awak:14,Mats:18}.
We furthermore expect more than 
one order of magnitude lower
background rates 
owing to the use of thinner (0.8~mm thick) 
CZT detectors, improved shielding, and flights closer
to solar minimum rather than solar maximum \citep[see][]{Shaw:03,Potg:08}.
Simulated {\it XL-Calibur} observations of GX 301$-$2 (Fig.\,\ref{f:xlc}) show that the improved 
mission could clearly distinguish 
between the fan beam and the pencil beam model.
Joint observations with the 
{\it Imaging X-Ray Polarimetry Explorer} ({\it IXPE}, 2\,keV-8\,keV, launch in 2021)  \citep[][]{Weis:16}
and {\it XL-Calibur} (launches in 2022, 2023, and 2025),
will enable detailed comparisons of 
predicted and observed signatures.
\section*{Appendix A - Stokes Parameter Analysis of the {\it X-Calibur} Data}
The analysis of the {\it X-Calibur} events starts  with the de-rotation of the $x$ and $y$ coordinates of the energy deposition in the detector reference frame into the reference frame of the telescope truss.
Subsequently, we correct for the offset of the focal 
spot of the X-ray mirror from the center of the scattering element 
as determined from the excess recorded in the rear CZT detector (Fig.\ \ref{f:17}). 
Finally, the coordinates are referenced to the celestial North pole 
based on the truss orientation measured by the pointing system.
Choosing a coordinate system with the 
$y$-coordinate pointing North and the 
$x$-coordinate pointing East, the 
azimuthal scattering angle 
is given by:
\begin{equation}
\psi=\arctan{(y/x)}
\end{equation}
so that $\psi=0$ corresponds to scatterings along the North-South direction, 
and $0<\psi<\pi/2$ corresponds to scatterings along the North-East direction.  
We calculate a set of Stokes parameters \citep{Kisl:15} for the 
\mbox{$k^{th}$}
event:
\begin{align}
  i_k   & = 1 \\
  q_k  & = -\frac{2}{\mu} \cos{(2\psi_k)}\\
  u_k  & = -\frac{2}{\mu} \sin{(2\psi_k)}
\end{align} 
The factor $\mu$ is the modulation factor (see Equation (\ref{e:mu})). The minus signs in the expressions of $q_k$ and $u_k$ account for the 90$^{\circ}$ offset between the electric field vector of the photons and the preferred scattering direction.The factor $2/\mu$ normalizes $q_k$ ($u_k$) so that its average is 1 for a beam 
100\% linearly polarized along the North-South direction (looking into the sky, 45$^{\circ}$ anti-clockwise from the North-South direction).

The $k^{\it th}$ event enters the analysis with weight $w_k$ that is proportional 
to the expected signal-to-background ratio, and is the 
product of two functions (spectral analysis)
or three functions (light curves) optimized with
Monte Carlo simulations of the detector.
The first function $f_1(z)$ depends on the position of the energy deposition along the optical axis (the $z$ coordinate) 
and accounts for the approximately exponential distribution of the depths of the Compton scattering in the 
scattering element. As a consequence, most source photons are detected near the front of the polarimeter.
The second function $f_2(x,y)$ depends on the position of the triggered pixel relative to the scattering element and
is proportional to the azimuthal scattering angle interval $\Delta \psi$ that the pixel covers as seen 
from the axis of the scattering element. 
The function weighs events close to the middle of the 
side walls of the rectangular detector assembly more 
heavily than those close to the edges, as those pixels
achieve a better signal-to-background ratio.
The third function $f_3(E)$ (only for light curves) weighs events
according to the energy $E$ deposited in the CZT detectors and is proportional to the expected source detection rate as a
function of energy accounting for the source spectrum, 
atmospheric absorption, and the mirror effective area.

With $t_{\rm ON}$ and $t_{\rm OFF}$ being the ON and OFF observation times and 
$\alpha =  t_{\rm ON}/t_{\rm OFF}$, we define the total background-subtracted Stokes 
parameters as:
\begin{align}
  I   & = \sum_{ON} w_k i_k- \alpha \sum_{OFF} w_k i_k\\    
  Q  & = \sum_{ON} w_k q_k - \alpha \sum_{OFF} w_k q_k\\
  U  & = \sum_{ON} w_k u_k - \alpha \sum_{OFF} w_k u_k,
\end{align}
where the sums run over the ON and OFF events.

Compared to the unweighted analysis, the weighted analysis improves the signal-to-background ratio  of the \mbox{GX 301$-$2} results by $\sim$20\%. 
Further sensitivity improvements might be achieved with a maximum likelihood
analysis \citep[see][]{Kraw:11b,Lowe:17a,Lowe:17b}. 

We calculate statistical errors on $I$, $Q$, $U$, ${\cal Q}$ and ${\cal U}$ from error propagation.
Each event contributes with the following RMS-values to the analysis \citep{Kisl:15}:
\begin{align}
 \sigma_{i_k}   & = 1 \\
  \sigma_{q_k}  & = \frac{\sqrt{2}}{\mu}\\
  \sigma_{u_k} & = \frac{\sqrt{2}}{\mu}
\end{align} 
The estimates of $\sigma_{q_k}$ and $\sigma_{u_k}$ are conservatively chosen for $p_0=0$.
For $p_0>0$ the errors are smaller. 
When calculating the error on ${\cal Q}$ (${\cal U}$), we assume that the errors on $I$ and $Q$
($I$ and $U$) are statistically independent. 
A toy simulation shows that this is indeed an
excellent assumption.
\section*{Appendix B - Systematic Errors on the {\it X-Calibur} Polarization Results}  
\label{sec:syst}
We calibrated the polarimeter at the Cornell High Energy Synchrotron Source using a 
40\,keV beam with a $\sim$90\% polarization \citep{Beil:14}. 
The measurements were 
carried out with different polarimeter orientations allowing
us to simulate an unpolarized beam by combining data taken at 
orientations differing by 90$^{\circ}$.
It is important to note that the rotation of the detector and shield assembly
removes systematic errors due to detector non-uniformities (e.g. dead pixels, noisy pixels) 
and geometrical effects 
(including uncertainties in the distances 
between the center of the scattering element and the CZT detectors and gaps between 
the detectors).
Based on the calibration data, we estimate that we know the modulation factor $\mu$ within an uncertainty of $\pm 2\%$. 
The uncertainty on $\mu$ introduces a relative systematic error on the 
measured polarization fraction $p_{\rm r}$ of $\Delta p_{\rm r}\,=\,2\% p_{\rm r}$.

The misalignment of the center of the mirror point-spread function and
the rotation axis of the polarimeter can lead to a spurious polarization
which is independent of the true polarization fraction \citep{Beil:14}. 
Based on the image of GX 301$-$2 in the rear CZT detector (Fig.\ \ref{f:17}), 
we estimate that the center of the PSF and the rotation axis 
of the polarimeter were offset by ($d \,=\, 1.5$\,mm). 
Correcting for $d$, the uncertainty in $d$ leads to a residual
systematic polarization fraction error of $<0.25\%$.

We performed the full Stokes analysis for the background data runs, 
and obtain Stokes parameters which are consistent with 0. For example, for the entire \mbox{15-35\,keV} background, we get:
\begin{align}
{\cal Q}_{\rm OFF} &= -0.015\pm0.011\\
{\cal U}_{\rm OFF} &= 0.010\pm0.011
\end{align}
where the errors are given for a 1\,$\sigma$ confidence interval (see also Fig.\,\ref{f:qu}). 
The fact that the background looks unpolarized implies that an under- or over-subtraction
of the background (owing for example to a time variable background) 
does not create a spurious polarization detection. 
We estimate that the background subtraction 
introduces a relative 5\% error on measured polarization fractions.  

Adding all systematic errors linearly, we get a total systematic error on the polarization fraction quoted in Equation (\ref{e:se}).
\section*{Appendix C - Data Tables}
\vspace*{-2ex}
\begin{center}
\begin{deluxetable*}{ccccc}[ht]
\tabletypesize{\scriptsize}
\tablecaption{Summary of {\it X-Calibur}, {\it NICER} and {\it Swift}-XRT observations. \label{tabO}}
\tablewidth{0.pt}
\tablehead{
\colhead{Instrument} & \colhead{Label} & \colhead{ObsID} &  \colhead{Start [MJD]} & \colhead{Exposure [s]}}
\startdata
{\it X-Calibur} & X-I & 1 & 58482.158310 & 1080\\ 
{\it X-Calibur} & X-II & 2 & 58482.168337 & 653\\ 
{\it X-Calibur} & X-III & 3 & 58482.188979 & 925\\ 
{\it X-Calibur} & X-IV & 4 & 58482.211214 & 923\\ 
{\it X-Calibur} & X-V & 5& 58482.233435 & 925\\ 
{\it X-Calibur} & X-VI & 6& 58482.255660 & 925\\ 
{\it X-Calibur} & X-VII & 7& 58482.277879 & 925\\ 
{\it X-Calibur} & X-VIII & 8& 58482.300110 & 924\\ 
{\it X-Calibur} & X-IX & 9& 58482.322312 & 928\\ 
{\it X-Calibur} & X-X & 10& 58482.344555 & 924\\ 
{\it X-Calibur} & X-XI & 11& 58482.366783 & 925\\ 
{\it X-Calibur} & X-XII & 12& 58482.389004 & 924\\ 
{\it X-Calibur} & X-XIII & 13& 58482.411225 & 925\\ 
{\it X-Calibur} & X-XIV & 14& 58482.433447 & 916\\ 
{\it X-Calibur} & X-XV & 15& 58482.455664 & 926\\ 
{\it X-Calibur} & X-XVI & 16& 58482.477913 & 923\\ 
{\it X-Calibur} & X-XVII & 17& 58482.500132 & 923\\ 
{\it X-Calibur} & X-XVIII & 18& 58482.522364 & 922\\ 
{\it X-Calibur} & X-XIX & 19& 58482.544571 & 926\\ 
{\it X-Calibur} & X-XX & 20& 58482.566831 & 918\\ 
{\it X-Calibur} & X-XXI & 21& 58482.589033 & 928\\ 
{\it X-Calibur} & X-XXII & 22& 58483.117441 & 219\\ 
{\it X-Calibur} & X-XXIII & 23& 58483.135091 & 757\\ 
{\it X-Calibur} & X-XXIV & 24& 58483.151973 & 986\\ 
{\it X-Calibur} & X-XXV & 25& 58483.174523 & 931\\ 
{\it X-Calibur} & X-XXVI & 26& 58483.193260 & 328\\ 
{\it X-Calibur} & X-XXVII & 27& 58483.218975 & 932\\ 
{\it X-Calibur} & X-XXVIII & 28& 58483.241197 & 930\\ 
{\it X-Calibur} & X-XXIX & 29& 58483.263374 & 925\\ 
{\it X-Calibur} & X-XXX & 30& 58483.285587 & 925\\ 
{\it X-Calibur} & X-XXXI & 31& 58483.307799 & 925\\ 
{\it X-Calibur} & X-XXXII & 32& 58483.330011 & 925\\ 
{\it X-Calibur} & X-XXXIII & 33& 58483.352284 & 936\\ 
{\it X-Calibur} & X-XXXIV & 34& 58483.374538 & 931\\ 
\hline
{\it NICER} & N-I & 1010220101 & 58,480.09 & 400\\ 
{\it NICER} & N-II & 1010220101 & 58,480.16 & 230\\ 
{\it NICER} & N-II & 1010220101 & 58,480.28 & 310\\ 
{\it NICER} & N-IV & 1010220101 & 58,480.34 & 1015\\ 
{\it NICER} & N-V & 1010220102 & 58,481.26 & 230\\
\hline
{\it Swift}-XRT & S-I & 00031256019 & 58,480.10 & 1055\\
{\it Swift}-XRT & S-II & 00031256020 & 58,481.15 & 1010\\
{\it Swift}-XRT & S-III & 00031256021 & 58,482.73 & 960\\
{\it Swift}-XRT & S-IV & 00031256022 & 58,483.66 & 960\\
{\it Swift}-XRT & S-V & 00031256023 & 58,484.00 & 760\\
{\it Swift}-XRT & S-VI & 00031256024 & 58,485.52 & 895\\
{\it Swift}-XRT & S-VII & 00031256025 & 58,486.39 & 990\\
{\it Swift}-XRT & S-VIII & 00031256026 & 58,487.38 & 540\\
{\it Swift}-XRT & S-IX & 00031256027 & 58,488.51 & 920\\
\enddata
\end{deluxetable*}

\begin{deluxetable*}{cc}[ht]
\tabletypesize{\scriptsize}
\tablecaption{GX 301$-$2 phase model parameters used in this paper. \label{t:eph}}
\tablewidth{0.pt}
\tablehead{
\colhead{Parameter} & \colhead{Value}}
\startdata
$t_0$& 58477.024509 (MJD)\\
$\dot{\phi}$ &  126.350509 day$^{-1}$\\
$\ddot{\phi}$ &  0.0769078 day$^{-2}$\\
$\dddot{\phi}$ &  -0.00868925 day$^{-3}$\\
$\ddddot{\phi}$ & 0.001075897 day$^{-4}$
\enddata
\end{deluxetable*}
\vspace*{-2ex}

\begin{deluxetable*}{lccccc}[ht]
\tabletypesize{\scriptsize}
\tablecaption{Spectral results from {\it NICER} observations. The errors are on 1$\sigma$ confidence level.\label{tabN}}
\tablewidth{0.pt}
\tablehead{
\colhead{Observation:} & \colhead{N-I} & \colhead{N-II} & \colhead{N-III}& \colhead{N-IV}& \colhead{N-V}}
\startdata
$F_{\rm 2-10\,keV}$ [$\times 10^{-9}\,{\rm erg}\,{\rm cm}^{-2}\,{\rm s}^{-1}$] & $1.20 \pm 0.06$ & $0.78 \pm 0.10$ & $1.29 \pm 0.08$ & $1.21 \pm 0.02$ & $2.36 \pm 0.12$ \\
N$_{\rm H,1}$ [$10^{22}$\,cm$^{-2}$]& $78.1 \pm 2.4$ & $82.5 \pm 4.5$ & $80.3\pm2.7$ & $86.5\pm1.8$ & $52.2\pm2.7$\\
N$_{\rm H,2}$ [$10^{22}$\,cm$^{-2}$]& $2.7 \pm 1.0$ & $4.4 \pm 1.6$ & $4.5\pm1.8$ & $2.6\pm0.6$ & $0.05\pm0.83$\\
Cov. Frac. & $0.992 \pm 0.001$ & $0.986 \pm 0.003$ & $0.994\pm0.001$ & $0.990\pm0.001$ & $0.961\pm0.005$\\
PL$_\textrm{Norm}$ [cm$^{-2}$s$^{-1}$keV$^{-1}$] & $0.53\pm 0.10$ & $0.29 \pm 0.09$ & $0.60\pm0.13$ & $0.30\pm0.04$ & $0.09\pm0.02$\\
PL$_{\Gamma}$ &  $1.21 \pm 0.08$ & $1.09 \pm 0.14$ & $1.21\pm0.09$ & $0.85\pm0.06$ & $0.16\pm0.09$\\
Fe K$\alpha$\,A [$10^{-3}$\,s$^{-1}$cm$^{-2}$] & $3.4\pm0.3$ & $2.6\pm0.3$ & $5.3\pm0.5$& $4.3\pm0.2$ & $5.0\pm0.7$\\
Fe K$\alpha$\,E [keV] & $6.41\pm 0.01$ & $6.39 \pm 0.01$ & $6.40 \pm 0.01$ & $6.41\pm0.01$ & $6.41\pm0.01$\\
Fe K$\alpha\,\sigma$ [keV] & $0.02\pm0.02$ & $0.02$ & $0.060\pm0.011$ & $0.039\pm0.007$ & $0.068\pm0.018$\\
$\chi^{2}/{\rm NDF}$ & 0.97 & 1.00 & 0.93 & 1.14 & 1.06 \enddata
\end{deluxetable*}
\begin{deluxetable*}{lccccccccc}[ht]
\tabletypesize{\tiny}
\tablecaption{Spectral results from the {\it Swift XRT}. The errors are on 1$\sigma$ confidence level.\label{tabS}}
\tablewidth{0.pt}
\tablehead{
\colhead{Observation:} & \colhead{S-I} & \colhead{S-II} & \colhead{S-III}& \colhead{S-IV}& \colhead{S-V} & \colhead{S-VI} & \colhead{S-VII} & \colhead{S-VIII} & \colhead{S-IX}}
\startdata
$F_{\rm 2-10\,keV}$ [$10^{-9}\,{\rm erg}\,{\rm cm}^{-2}\,{\rm s}^{-1}$] & $1.15 \pm 0.20$ & $1.53 \pm 0.12$ & $2.27 \pm 0.15$ & $2.29 \pm 0.11$ & $3.69\pm0.15$ & $2.81 \pm 0.12$ & $1.91 \pm 0.09$ & $1.35\pm0.20$ & $1.16\pm0.12$ \\
N$_{\rm H,1}$ [$10^{22}$\,cm$^{-2}$]& $63.6\pm5.8$ & $76.5\pm12.3$ & $49.1\pm2.9$ & $67.6\pm8.0$ & $49.4\pm3.1$ & $32.8\pm1.4$ & $39.3\pm2.0$ & $42.4\pm5.2$ & $61.6\pm19.3$\\
N$_{\rm H,2}$ [$10^{22}$\,cm$^{-2}$]& $0.0$ & $21.1\pm10.1$ & $0.0$ & $16.3\pm9.8$ & $0.0$ & $0.0$ & $0.0$ & $0.0$ & $19.8\pm9.4$\\
Cov. Frac. & $1.0$ & $0.945\pm0.045$ & $1.0$ & $0.970\pm0.031$ & $0.988\pm0.002$ & $1.0$ & $1.0$ & $1.0$ & $0.858\pm0.134$\\
PL$_\textrm{Norm}$ [cm$^{-2}$s$^{-1}$keV$^{-1}$] & $0.32\pm0.17$ & $0.80\pm0.41$ & $0.23\pm0.07$ & $0.84\pm0.34$ & $0.33\pm0.09$ & $0.46\pm0.09$ & $0.25\pm0.06$ & $0.12\pm0.07$ & $0.38\pm0.25$\\
PL$_{\Gamma}$ & $1.91\pm0.36$ & $1.33\pm0.22$ & $1.09\pm0.24$ & $1.19\pm0.17$ & $0.61\pm0.12$ & $1.01\pm0.09$ & $0.85\pm0.12$ & $0.67\pm0.26$ & $1.23\pm0.27$\\
Fe K$\alpha$\,A [$10^{-3}$\,s$^{-1}$cm$^{-2}$] & $9.9\pm2.4$ & $5.0\pm1.8$ & $12.2\pm3.6$ & $7.6\pm2.0$ & $18.6\pm3.4$ & $8.9\pm2.8$ & $8.9\pm1.4$ & $9.5\pm6.4$ & $3.8\pm1.1$\\
Fe K$\alpha$\,E [keV] & $6.56\pm0.07$ & $6.34\pm0.05$ & $6.37\pm0.05$ & $6.49\pm0.04$ & $6.52\pm0.05$ & $6.47\pm0.04$ & $6.43\pm0.03$ & $6.33\pm0.06$ & $6.33\pm0.06$\\
Fe K$\alpha\,\sigma$ [keV] & $0.317\pm0.082$ & $0.144\pm0.085$ & $0.316\pm0.070$ & $0.127\pm0.073$ & $0.296\pm0.061$ & $0.203\pm0.058$ & $0.145\pm0.048$ & $0.301\pm0.078$ & $0.200$\\
${\chi}^2/{\rm NDF}$ & 1.58 & 1.53 & 1.07 & 0.90 & 1.28 & 1.13 & 0.75 & 0.66 & 1.20 
\enddata
\end{deluxetable*}
\end{center}
\clearpage
\section*{Acknowledgements}
We thank  A. Awaki (Ehime Uniersity), K. Hayashida (Osaka University, Project Research Center for Fundamental Sciences, ISAS), Y. Maeda (ISAS), H. Matsumoto (Osaka University, Project Research Center for Fundamental Sciences), T. Tamagawa (RIKEN), and K. Tamura (Nagoya University) for fruitful discussions and for their comments on this paper. We thank V. Mikhalev for contributing the code for barycentring the {\it X-Calibur} event times and Rakhee Kushwah (KTH, Oskar Klein Centre) for contributing to the flight monitoring shifts.

{\it X-Calibur} is funded by the NASA APRA program under contract number 80NSSC18K0264. We thank the McDonnell Center for the Space Sciences at Washington University in St. Louis for funding of an early polarimeter prototype, as well as for funds for the development of the ASIC readout.   
Henric Krawczynski acknowledges NASA support under grants 80NSSC18K0264 and NNX16AC42G.
KTH authors acknowledge support from the Swedish National Space Agency (grant number 199/18). MP also acknowledges support from the Swedish Research Council (grant number 2016-04929).
Hans Krimm acknowledges support from the National Science Foundation under the Independent Research and Development program.

\end{document}